\documentclass[12pt]{iopart}

\usepackage{hyperref}
\usepackage{amssymb}   
\usepackage{xcolor}
\usepackage{graphicx}
\usepackage[numbers,sort&compress]{natbib}

\newcommand{\orcid}[1]{\href{https://orcid.org/#1}{ORCID}}

\newcommand{\rr}{\vec{r}}
\newcommand{\RR}{\vec{R}}

\newcommand{\rrho}{\vec{\rho}}

\begin{document}

\title{Moir\'e excitons and exciton-polaritons: A review}

\author{Sa\'ul A. Herrera-Gonz\'alez \orcid{0000-0001-7404-9630},
Hugo A. Lara-Garc\'ia \orcid{0000-0002-3348-1133},
Giuseppe Pirruccio \orcid{0000-0001-7848-2853},
David A.\ Ruiz-Tijerina \orcid{0000-0001-7688-2511},
Arturo Camacho-Guardian \orcid{0000-0001-5161-5468}}

\address{Departamento de F\'isica Qu\'imica, Instituto de F\'isica, Universidad Nacional Aut\'onoma de M\'exico, Ciudad de M\'exico, C.P.\ 04510, M\'exico}
\ead{ saulaherrera@fisica.unam.mx,acamacho@fisica.unam.mx}

\vspace{10pt}
\begin{indented}
\item[]June 2025
\end{indented}

\begin{abstract}
Distinguished by their long lifetimes, strong dipolar interactions, and periodic confinement, moiré excitons are fertile ground for realizing interaction-driven excitonic phases beyond conventional semiconductor systems. Formed in twisted or lattice-mismatched van der Waals heterostructures, these excitons are shaped by a periodic potential landscape that enables the engineering of flat bands, strong interactions, and long-lived localised states. This has opened pathways to explore strongly correlated phases, including excitonic insulators, superfluids, and supersolids, potentially stable even at room temperature. When embedded in optical cavities, moiré excitons hybridize with photons to form moiré exciton-polaritons, a new class of quasiparticles exhibiting enhanced optical nonlinearities and novel topological features. In this review, we survey the theoretical foundations and experimental progress in the field of moiré excitons and polaritons. We begin by introducing the formation mechanisms of moiré patterns in two-dimensional semiconductors, and describe their impact on exciton confinement, optical selection rules, and spin-valley physics. We then discuss recent advances in the realization of many-body excitonic phases and exciton-based probes of electronic correlations. Finally, we explore the novel aspects of moiré polaritons, highlighting their unique nonlinear and topological properties. By bridging quantum optics, nanophotonics, and correlated electron systems, moiré excitons offer a powerful solid-state platform for quantum simulation, optoelectronic applications, and many-body photonics.
\end{abstract}

\section{Introduction}

The arrival of atomically thin materials has revolutionised condensed matter physics and material science. Two-dimensional (2D) van der Waals (vdW) materials like graphene ~\cite{Geim2009grapheneREVIEW}, transition metal dichalcogenides (TMDs) ~\cite{Mak2016theREVIEW}, and hexagonal boron nitride (hBN) ~\cite{Caldwell2019photonicsREVIEW} have spurred intense research efforts in view of their unique electronic, optical, and mechanical properties. In 2D semiconductors, the reduced dimensionality tends to enhance Coulomb interactions between their charge carriers, enabling strong light-matter coupling and making them ideal platforms to realize unexplored quantum many-body phenomena. The ability to tune the electronic and optical properties of these systems with an unprecedented degree of control has also placed 2D materials as ideal candidates for quantum simulators ~\cite{Wu2018Hubbard,Tang2020simulation,Kennes2021moire}, as well as for the development of new optoelectronic devices~\cite{Zhang2016vanREVIEW,Mak2016theREVIEW,He2021moireREVIEW,Guo2020stackingREVIEW,Du2023moireREVIEW}.

A powerful idea from back in the early 2010s is the fabrication of {\it van der Waals heterostructures}, where multiple monolayers are vertically stacked to engineer a material with new physical properties~\cite{Geim2013fabrication,NovoselovScience2016,Guo2020stackingREVIEW}. This possibility, afforded by the 2D nature of vdW materials, can lead to an {\it emergent crystal} with a lattice structure determined by the particular choice of stacking arrangement.
In particular, when there is a small lattice mismatch or twist angle between the stacked layers, a long-wavelength interference pattern known as a {\it moir\'e superlattice} arises \cite{Li2010observation,Luican2011single,He2021moireREVIEW}. The emergent pattern defines a new potential for electrons, holes, and excitons, dramatically modifying the particle and quasiparticle band structure, as well as their interactions \cite{Santos2007graphene,Morell2010flat,Bistritzer2011moire,Wu2017topological,Yu2017moire,Wu2018Hubbard}. 

Moir\'e superlattices have unveiled a rich landscape of electronic phenomena, ranging from the engineering and control of electronic band structures, such as the formation of flat bands, to the realization of exotic strongly correlated states, including unconventional superconductivity \cite{Cao2018unconventional,Lu2019superconductors,Balents2020superconductivityREVIEW}, correlated insulators \cite{Cao2019correlated,Tang2020simulation,Mak2022semiconductorREVIEW} and topological phases of matter~\cite{Cai2023signatures,Morales2025fractionalized}. 

The moir\'e pattern introduces nanoscale periodic confinement for excitons---Coulomb bound electron-hole pairs---which can become localised at the potential minima across the superlattice \cite{Yu2017moire,Seyler2019signatures}. Depending on the specific stacking configuration, excitons may localise into quantum-dot-like states or partially delocalise into minibands \cite{Yu2017moire,Wu2018theory,Ruiz2020theory,Magorrian2021multifaceted}. These {\it moir\'e excitons} exhibit tunable binding energies, optical selection rules, spin-valley dynamics, transport properties and light-matter coupling, and possess long lifetimes, making them a promising platform for realizing rich many-body phenomena such as excitonic insulators, superfluids, and Wigner crystals~\cite{Wu2018Hubbard,Kennes2021moire,Mak2022semiconductorREVIEW,Regan2022emerging}. Their sensitivity to stacking arrangement, twist angles between layers, dielectric environment and external fields~\cite{Yu2017moire,Wu2018Hubbard,Tang2020simulation,Torre2024advancedREVIEW,He2021moireREVIEW} enables external control over these correlated excitonic phases.

The interaction between moir\'e excitons and light can be enhanced by embedding a vdW heterostructure into a photonic cavity \cite{Dufferwiel2017,Dufferwiel2018,Emmanuele2020,LaMountain2021,Zhang2021van}. In this configuration the strong coupling between excitons and cavity photons gives rise to hybrid light-matter quasiparticles known as \textit{exciton-polaritons}. 
Moir\'e exciton-polaritons inherit characteristics of the moir\'e superlattice through their excitonic component \cite{Fitzgerald2022}, resulting in a new class of polaritons with properties that markedly differ from those obtained with conventional 2D semiconductors. These include novel strong nonlinearities, topological polaritonic states, and the potential realization of quantum many-body phases of light.

In this Review we aim to provide a comprehensive overview of the progress and state of the art of moir\'e excitons and polaritons. We survey key experimental and theoretical breakthroughs, highlighting recent progress in the physics of many-body excitonic phases, and their strong light-matter coupling.

This field stands at the intersection of quantum optics, nanophotonics, and strongly correlated matter. By combining the design freedom of vdW assembly with precise optical control, moiré exciton and polariton systems are an exciting frontier for both fundamental physics and device applications, including quantum information processing, ultra-efficient light sources and valleytronic components \cite{Zhang2016vanREVIEW,Vitale2018valleytronicsREVIEW,Mak2018lightREVIEW,Liu2019valleytronicsREVIEW,An2021perspectivesREVIEW,He2021moireREVIEW,Wang2021stackingREVIEW,Abbas2020recentREVIEW,Tran2020moireREVIEW,Torre2024advancedREVIEW,Du2023moireREVIEW}.

This Review is organized as follows. In Section \ref{sec:Xin2DSemic}, we provide an overview of excitons in two-dimensional semiconductors, including their binding mechanisms, spin-valley properties, and sensitivity to external fields. Section~3 introduces moiré patterns in TMD heterostructures and discusses the resulting moiré excitons from both theoretical and experimental perspectives. Section~4 focuses on quantum many-body phenomena enabled by moiré-confined excitons and exciton-carrier (Bose-Fermi) mixtures, such as Mott insulating states and superfluidity. 
In Section~5, we discuss the formation and properties of moiré exciton-polaritons, emphasizing recent advances in nonlinear optics, topological effects, and driven-dissipative quantum phases. We conclude with an outlook on open questions and emerging directions in this rapidly evolving field.

\section{Excitons in 2D semiconductors}\label{sec:Xin2DSemic}
\subsection{Screened interactions and the Wannier-Mott equation}\label{sec:WannierMottEq}

In 2D materials, Coulomb interactions between charge carriers (electrons and holes) are typically enhanced, compared with their 3D bulk counterparts \cite{Wang2018colloquiumREVIEW}. This enhancement arises from a combination of reduced dielectric screening in atomically thin materials, and the confinement of the charge carriers to a single plane. In contrast to bulk systems, where screening is governed by the three-dimensional dielectric function, the effective screening in two-dimensional materials is non-local and determined by both the intrinsic dielectric properties of the layer and the dielectric environment of the surrounding materials. As a consequence, excitons, bound electron-hole pairs, form with large binding energies, typically in the range of hundreds of meV\cite{Wang2018colloquiumREVIEW}. These values are one to two orders of magnitude larger than in conventional bulk semiconductors, such as GaAs\cite{XinGaAs} or Si, making the excitons stable even at room temperature~\cite{Mueller2018excitonREVIEW,Sun2022enhanced, Regan2022emerging}.

The strong binding of excitons in 2D semiconductors can give rise to a large oscillator strength, enabling efficient coupling to light despite the atomic-scale thickness of these materials. This 
leads to pronounced absorption and emission features, which dominate the optical response of the material, 
even at ambient conditions. This makes 2D semiconductors prime candidates for the exploration of light-matter interactions and exciton-based optoelectronics~\cite{Bhimanapati2015recentREVIEW,Wang2018colloquiumREVIEW,Du2023moireREVIEW}.

The binding of an electron-hole pair leading to the formation of an exciton in 2D semiconductors can be described by the Wannier-Mott equation\cite{WannierEq}: an effective-mass two-body Hamiltonian, which in terms of the relative vector $\rrho$  and center-of-mass (COM) vector $\RR$ typically reads~\cite{Wang2018colloquiumREVIEW, Lu2019modulated,Ruiz2020theory}:
\begin{equation}\label{eq:wanniermott}
    H_{e\textrm{-}h} = \frac{P^2}{2M} + \frac{p^2}{2\mu} + U_K(\rrho),
\end{equation}
where $\vec{P}$ and $\vec{p}$ are the COM and relative motion (RM) momenta, $M = m_e + m_h$ is the total exciton mass, and $\mu = \frac{m_e m_h}{M}$ is the reduced mass of the exciton. The term $U_K$ describes the electron–hole interaction potential, consisting of their mutual Coulomb attraction, screened by the material itself, as well as its dielectric environment.

Interestingly, unlike in 3D systems where the Coulomb interaction follows the familiar $1/\rho$ form, in 2D semiconductors the screening effects result in a modified potential, commonly modelled~\cite{Cudazzo2011dielectric,Wang2018colloquiumREVIEW,Ruiz2020theory} by the Rytova-Keldysh form~\cite{rytova2018screened,Keldysh1979coulomb}
\begin{equation}\label{eq:keldysh}
U_K(\rho) = -\frac{\pi e^2}{2\epsilon r_{\mathrm{eff}}} \left[ H_0\left( \frac{\rho}{r_{\mathrm{eff}}} \right) - Y_0\left( \frac{\rho}{r_{\mathrm{eff}}} \right) \right],
\end{equation}
where $H_0$ and $Y_0$ are a Struve function and a Bessel function of the second kind, respectively, and $r_{\mathrm{eff}}$ is the effective screening length, proportional to the ratio of the 2D material’s in-plane polarizability and the effective dielectric constant of its environment~\cite{Cudazzo2011dielectric}.
This potential interpolates between a logarithmic, 2D-like interaction at short distances, and a 3D Coulomb tail at long distances, capturing the essential features of electrostatics for a 2D material embedded in 3D space.

\begin{figure}[h]
\centering
\includegraphics[width=0.5\linewidth]{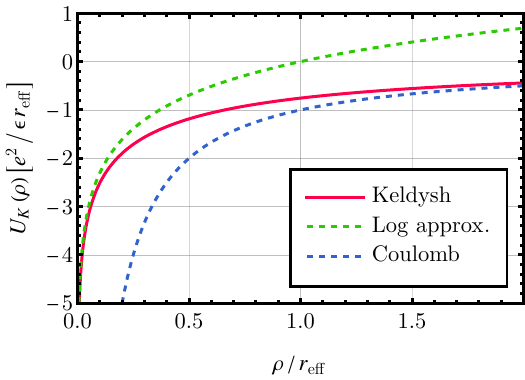}
\caption{
The interparticle electrostatic potential in van der Waals materials is described by the Keldysh (or Rytova-Keldysh) potential $U_K(\rho)$ (plotted as a function of normalized distance $\rho/r_{\mathrm{eff}}$, with $r_\mathrm{eff}$ the effective screening length). The Keldysh potential exhibits a crossover, from logarithmic behavior at short distances, to a $1/\rho$ (Coulomb) decay at long distances.}
\label{fig:keldysh_potential}
\end{figure}

The Wannier-Mott equation \eref{eq:wanniermott} with the Rytova-Keldysh interaction \eref{eq:keldysh} is commonly solved numerically with variational\cite{ChernikovPRL2014}, finite elements\cite{Danovich2018}, or direct diagonalisation methods\cite{Ruiz2020theory,Ceferino2020}. Calculations reveal a nonhydrogenic series of exciton energies, in excellent agreement with experimental values for up to the $5s$ exciton energy in monolayer TMDs\cite{ChernikovPRL2014} from reflectance contrast measurements.

\subsection{Intralayer and interlayer excitons}

\begin{figure}[t]
\centering
\includegraphics[width=0.8\linewidth]{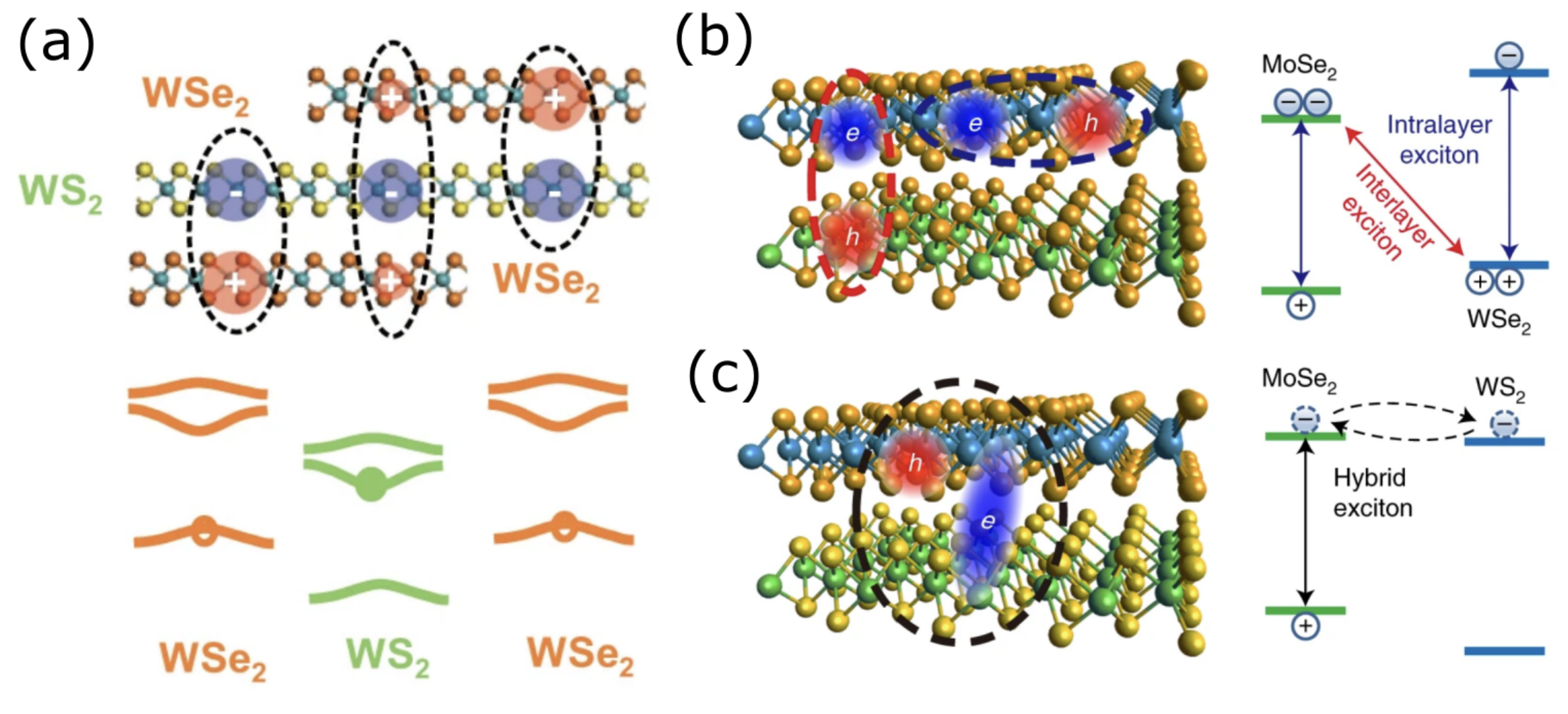}
\caption{\textbf{Dipolar, hybrid, and quadrupolar excitons in van der Waals heterostructures.}
\textbf{(a)} In structures such as MoSe$_2$/WS$_2$ heterostructures, the hybridization of electronic states across layers can lead to hybrid excitons, with mixed character from both monolayers. In trilayer moiré structures like WSe$_2$/WS$_2$/WSe$_2$, two interlayer excitons with opposite dipole moments can hybridize into a \emph{quadrupolar exciton}, which carries no net dipole but a finite quadrupole moment. These quadrupolar excitons exhibit quadratic Stark shifts and reduced exciton-exciton interactions compared to dipolar excitons, image adapted from \emph{Lian et al.} ~ \cite{Lian2023quadrupolar}.
\textbf{(b)-(c)} In a MoSe$_2$/WSe$_2$ heterobilayer with type-II band alignment, both intralayer and interlayer excitons can form. Intralayer excitons have electron and hole confined within the same monolayer, while interlayer (dipolar) excitons consist of spatially separated electrons and holes residing in adjacent layers, exhibiting a built-in out-of-plane dipole moment. Electron tunneling allows for the hybridisation of intra- and inter-layer excitons. Image adapted from \emph{Huang et al.} ~\cite{Huang2022excitonsREVIEW}.
}
\label{fig:exciton_types}
\end{figure}

Beyond the excitonic states intrinsic to monolayers of 2D semiconductors, the stacking of multiple layers results in a whole new class of exciton, whose electron and hole can reside in different layers: the \textit{interlayer} exciton (IX) \cite{Rivera2018interlayerREVIEW}. Interlayer excitons can exhibit a wide range of behaviours depending on the specific materials. However, the most studied ones are dipolar interlayer excitons that occur in simple bilayer systems, whether the layers are made of the same material (homo-bilayers) or different materials (hetero-bilayers). Recently, so-called quadrupolar IXs lacking a permanent electric dipole have been predicted and observed in symmetric trilayers~\cite{Bai2023evidence,Yu2023observation,Li2023quadrupolar,Deilmann2024}. These distinct types of IXs are illustrated in Fig.~\ref{fig:exciton_types}(a).

For IXs, the spatial separation between charge carriers leads to remarkable differences compared to intralayer excitons.  First, it yields an intrinsic electric dipole moment oriented perpendicular to the layers, rendering the exciton energy tunable via an applied out-of-plane electric field \cite{Yu2017moire,Tang2021tuning,Barre2022optical,Kim2025moire,Huang2025electrically,Karni2019infrared,Zeng2020observation,Joshi2020localized,Peng2022spatially}. Second, the suppression of interlayer recombination pathways by the permanent separation of the electron and hole significantly increases the IX lifetime, often by several orders of magnitude, as compared to its intralayer counterpart~\cite{Rivera2015observation,Choi2021twist,Miller2017,Xu2024control,Alexeev2024nature,Li2021interlayer,Mahdikhanysarvejahany2021temperature,Li2021dielectric}. In parallel, interlayer excitons also display remarkable transport properties, with large diffusion coefficients and extended propagation lengths reported in experiments~\cite{Fowler2024transport,Arora2021interlayer,Wang2021diffusivity,Bange2023ultrafast}. Finally, the out-of-plane permanent dipole promotes stronger electrostatic interactions between IXs~\cite{Kremse2020discrete,Debnath2022}. 
These enhanced lifetimes and interactions make IXs particularly attractive to realize systems with long-range coherence~\cite{Paik2019interlayer}, exploring many-body bosonic phenomena~\cite{Dutta2015nonstandardREVIEW}, such as exciton Bose–Einstein condensation~\cite{Eisenstein2004bose,Wu2015theory,Wang2019evidence,Guo2022tuning}, different types of excitonic insulators \cite{Jerome1967excitonic,Hu2018fractional,Xie2023nematic}, superfluidity~\cite{Fisher1989boson,Greiner2002quantum} and supersolidity~\cite{Iskin2011route,Zhang2021su4}. However, the increased lifetime of IXs comes at the price of strongly suppressed optical activity \cite{AnomalousLightCones2015,Rivera2015observation,Rivera2016valley,Zhang2019highly}. The reduced wavefunction overlap between the vertically separated electron and hole results in a small oscillator strength, limiting their ability to couple efficiently with light~\cite{Ross2017interlayer}. Moreover, in misaligned  structures the relative twist between the layers results in a mismatch between the electron and hole states in reciprocal space. This momentum mismatch gives the resulting IX a finite momentum in its ground state, suppressing its interaction with light, particularly at large twist angles~\cite{Forg2019,Choi2021twist}.

A particularly rich regime arises when interlayer and intralayer excitons coexist and hybridise. Hybridisation becomes significant when the conduction or valence band edges of neighboring layers align closely in energy and momentum~\cite{Alexeev2019resonantly,Ruiz2019interlayer}. This situation is common in homobilayer systems such as WSe$_2$/WSe$_2$ \cite{Brem2020hybridized,Scuri2020electrically,Merkl2020twist} and MoSe$_2$/MoSe$_2$ \cite{Sung2020broken}, but it can also arise in certain heterobilayers, depending on the band alignment and twist angle ~\cite{Alexeev2019resonantly,Hsu2019tailoring,Shimazaki2020strongly,Tang2021tuning,Chang2023,Polovnikov2024field,Zhao2024hybrid}. The resulting hybrid excitons then inherit features of both intra and interlayer excitons: a sizable oscillator strength from their intralayer component, which enables strong coupling to optical modes, and a permanent dipole moment from the interlayer component, which promotes strong exciton-exciton interactions. Thus, hybrid excitons are an ideal platform for studying strongly interacting polaritons, especially when embedded in photonic cavities or optical micro structures (see Sec.~\ref{Sec: polaritons}). Furthermore, their properties such as energy dispersion, optical selection rules, and spatial localisation can be substantially modified by the presence of a moir\'e potential, which we explore in the following sections.

It is worthwhile noting that exciton hybridisation does not require a band description, and can also occur at the level of excitonic wavefunctions, for instance, involving both Wannier-Mott excitons in a solid and Frenkel excitons in a molecule.  This scenario has been realised in organic–TMD heterostructures, where it leads to the formation of hybrid Frenkel–Wannier excitons ~\cite{Agranovich1998,Fu2024}, which also combine properties of both of its components. Notably, these hybrid excitons exhibit enhanced oscillator strength, inherited from their strongly localised Frenkel component.

\subsection{Spin-valley properties}

TMDs became highly attractive in view of their spin-valley properties, which inspired the field of {\it valleytronics}~\cite{Rivera2016valley,Vitale2018valleytronicsREVIEW,Mak2018lightREVIEW,Zhang2019highly,Unuchek2019valley,Liu2019valleytronicsREVIEW,Wang2019gigant}. For TMDs, their potential for valleytronics is strongly tied to their so-called spin-valley locking and their valley-specific optical selection rules~\cite{XiaoSpinValley2012}, which together allow for the control and readout of valley and spin degrees of freedom using polarised light. In typical TMDs, the direct band gap at the $K$ and $K'$ points of the Brillouin zone is spin split due to the strong spin-orbit coupling (SOC), especially significant in the valence band, leading to two excitonic resonances in the optical spectrum referred to as $A$ and $B$ excitons ~\cite{Mak2010Atomically}. The spin splitting of the conduction band plays an important role in determining the fine details of the optical properties. Particularly, molybdenum-based (MoX$_2$) and tungsten-based (WX$_2$) compounds exhibit opposite SOC splitting in their conduction bands \cite{Kormanyos2015kp}, with the minimum energy gap occurring for bands of opposite (same) electron spin in WX$_2$ (MoX$_2$) monolayers. As a result, the lowest energy excitons are optically bright in MoX$_2$, but optically dark in WX$_2$ \cite{Zhang2015experimental}. Thus, the nature of the dominant spin and valley transitions and consequently the optical selection rules depend sensitively on the chemical composition of the TMD.

The reduced dimensionality of the electron-hole Coulomb exchange interaction in TMDs leads to a light-like linear spectrum for excitons \cite{Yu2014Dirac} which was probed only recently \cite{Liu2025Direct}. Exchange also introduces intervalley scattering, leading to valley depolarization of excitons \cite{Yu2014Valley,Glazov2014Exciton, Qiu2015Nonanalyticity}. In moir\'e heterostructures, exchange can lead to a large Zeeman splitting \cite{Gomez2022Optical,Zhu2024Moire} and to F\"orster coupling \cite{Li2024Cross,Zheng2025Forster}.

When multiple TMD layers are stacked, the heterostructure optical selection rules can differ strongly from those of the individual layers, depending on the twist angle\cite{Choi2021twist}, stacking order \cite{Lian2024valley,Lou2024stacking} and number of layers \cite{Li2023quadrupolar,Lian2023quadrupolar,Yu2023observation}, among other factors~\cite{Yu2017moire,Regan2022emerging,Kim2024indentification}. 
As a result, the optical properties of multilayer TMD heterostructures become highly tunable, enabling dynamic control of valley physics via external fields, strain, and interlayer biasing~\cite{Mak2016theREVIEW,Vitale2018valleytronicsREVIEW,Liu2019valleytronicsREVIEW,Shinokita2022valley,Ciarrocchi2019polarization,Chen2025strain}. In what follows, we focus on how these moiré superlattices reshape the exciton landscape of TMDs and enable novel regimes of quantum light–matter interaction.

\subsection{Optical Spectrum Engineering: External Fields, Strain, Pressure, and Alloying}\label{sec:OptSpecEng}

2D materials offer a unique platform where external perturbations such as electric fields \cite{Rivera2015observation,Yu2017moire,Tang2021tuning,Ghiotto2021quantum,Li2021continuous,Hagel2022electrical,Dandu2022electrically,Lian2023quadrupolar,Tan2023layer,Chatterjee2023harmonic,Chen2023electrical,Tagarelli2023electrical,Xu2025valley,Knuppel2025correlated,Huang2025electrically,Kwak2025electrically,Zhang2025moire}, strain \cite{He2013experimental,Island2016precise,Yu2017moire,Bai2020excitons,Zhao2021dynamic,Hu2023moire,Wu2024optical,He2024unveiling,Kai2025distinct,Jiang2025anisotropic}, pressure \cite{Yankowitz2019tuning,Song2019switching,Li2022dynamic,Xie2023pressure,Xie2025pressure,Yan2025finely} and dielectric environment \cite{Stier2016probing,Raja2017coulomb,Xu2021creation,Sun2022enhanced} can directly exert an influence on the electronic and excitonic properties at the microscopic level. This sensitivity enables precise control over parameters such as the bandgap, exciton binding energy and interlayer coupling, making TMDs highly tunable potential platforms for both fundamental and technological applications.

Electrostatic gating provides control over the carrier density in two-dimensional semiconductors by injecting electrons or holes, effectively shifting the Fermi level. This not only modifies the free carrier population, but also screens Coulomb interactions, thereby renormalising exciton binding energies and the quasiparticle bandgap. Importantly, the presence of excess carriers can give rise to  few-body bound states such as trions: charged excitonic complexes formed by the binding of an exciton with an additional electron or hole \cite{Mak2013tightly,Brotons2021moire,Baek2021optical,Liu2021signatures,Chen2023excitonicREVIEW}. These few-body excitonic complexes exhibit distinct optical signatures and play a key role in the gate-tunable photophysics of 2D materials \cite{Sun2024dipolarREVIEW,Choi2024emergenceREVIEW}.

Strain engineering represents a different approach to modulating the excitonic properties of 2D semiconductors~\cite{Manzeli20162dREVIEW}. It exploits the high sensitivity of the electronic band structure to lattice deformations where, for instance, uniaxial or biaxial strain can not only significantly change the energies of exciton resonances \cite{He2013experimental,Island2016precise,Zhao2021dynamic,Jiang2025anisotropic}, but also modify the lattice symmetry, and thus the optical selection rules. This can induce an anisotropic optical response, changing the photoluminescence emission from circular to linearly polarized \cite{Bai2020excitons}. Along the same lines,  hydrostatic pressure has been shown to tune exciton binding energies and enhance interlayer coupling by reducing the interlayer distance~\cite{Pimenta2023pressure,Xie2024long}, a strategy widely employed in graphene-based systems~\cite{Yankowitz2018}. Finally, the ultrafast dynamics between excitons  allows us to measure the binding energy of interlayer excitons~\cite{Merkl2019ultrafast}.

The dielectric environment of 2D materials also plays a pivotal role in engineering excitonic properties, and has even been used to realize periodic potentials in monolayers~\cite{Forsythe2018band,Xu2021creation}. Substrates and encapsulating layers modify the effective screening of Coulomb interactions~\cite{Raja2017coulomb,Sun2022enhanced}, thereby tuning the exciton binding energy and radiative lifetime~\cite{Lin2014dielectric,Tan2023layer}. As for chemical methods, alloying TMD monolayers (e.g., Mo$_x$W$_{1-x}$S$_2$) enables the engineering of band gaps and excitonic resonances in a controllable manner, creating new opportunities for band structure tailoring\cite{alloys1,alloys2}.

In 2D heterostructures, these tuning strategies coexist with the formation of  moiré superlattices, resulting in a rich scenario where the underlying moiré potential, character of the optical excitations, light-matter coupling, interaction between the excitons, and the formation of trions and other excitonic complexes, etc., can be tuned on demand. This remarkable level of control has lead to the observation of excitonic insulators, Hubbard-like physics, novel polariton states, and other many-body excitonic phenomena that we will discuss in this Review.

\section{Moir\'e excitons}\label{sec:moireexcitons}

In stacked 2D semiconductors, moiré patterns arise from a lattice mismatch, a finite twist angle between layers, or both. These patterns correspond to spatial modulations of the atomic registry between the layers, which in turn modify the local point symmetry, interlayer distance, and consequently the local electronic properties. As a result, excitons perceive the moir\'e pattern as a spatially periodic potential landscape, with barriers that repel them and wells that can trap them, thus reshaping their energy spectra. The precise nature of these \emph{moir\'e excitons}, first reported in~\cite{Zhang2018moire,Seyler2019signatures,Alexeev2019resonantly,Jin2019observation,Tran2019evidence}, is determined by the underlying moir\'e potential, and as such they are susceptible to manipulation through direct control of the moir\'e pattern by any of the means discussed in Sec.\ \ref{sec:OptSpecEng}.

We begin by discussing moiré excitons in the linear (low-density) regime, where interactions between excitons and saturation effects can be neglected. In this regime, the excitonic features are well described by single-particle or non-interacting models.

\subsection{Moir\'e patterns in 2D semiconducting heterostructures}

Long-range periodic patterns emerge in stacks of 2D materials when there is a small lattice mismatch $\delta\ll 1$ between the layers, or a small relative twist angle (in radians) $\theta\ll1$ between their crystallographic axes, which lead to spatially varying local atomic registries. This modulation forms the moiré pattern: an interference effect rooted in the superposition of two slightly mismatched periodic lattices. Indeed, moiré patterns occur widely in optics, crystallography,  acoustics, and several other fields in physics~\cite{Du2023moireREVIEW,amidror2009theory,patorski1993handbook,yokozeki1976geometric,theocaris1972vibration}. For electrons, the moir\'e pattern translates into a periodic modulation of the local electrostatic potentials and interlayer couplings, resulting in what is commonly known as the moir\'e potential.

\begin{figure}[t]
\centering
\includegraphics[width=0.85\linewidth]{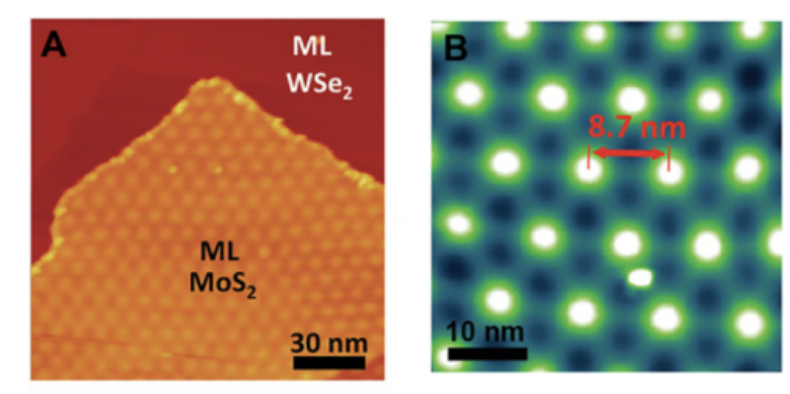}
\caption{\textbf{Moiré superlattice in a rotationally aligned MoS$_2$/WSe$_2$ hetero-bilayer.}
\textbf{(A)} Scanning tunneling microscopy (STM) image of a directly grown monolayer (ML) MoS$_2$/WSe$_2$ heterostructure, showing clear atomic resolution and long-range moiré modulation due to the lattice mismatch $\delta \approx 0.036$. 
\textbf{(B)} Zoom-in STM image highlighting the hexagonal moiré pattern with a measured periodicity of 8.7 nm, corresponding to the interference of the lattice constants of MoS$_2$ (3.16 Å) and WSe$_2$ (3.28 Å) under R-type stacking.
The resulting moiré potential gives rise to a lateral modulation of local electronic structure, forming a 2D electronic superlattice with site-dependent band edges and local bandgaps. This leads to spatially modulated interlayer exciton energies and offers a platform for exploring exciton confinement and quantum dot-like behavior in moiré traps. Figures adapted from \emph{Zhang et al.}~\cite{Zhang2017}.
}
\label{fig:moire_hetero}
\end{figure}

The moiré periodicity $l$ of a bilayer formed by two 2D crystals with lattice constants $a_0$ and $a_1 = a_0(1+\delta)>a_0$ (with dimensionless lattice mismatch $\delta = (a_1 - a_0)/a_0$) and an interlayer twist angle $\theta$ between the layer, is approximately given by~\cite{Mak2022semiconductorREVIEW}
\begin{equation}
l \approx \frac{a_0}{\sqrt{\delta^2 + \theta^2}},
\end{equation}
implying that $l$ can take values much larger than that of the atomic lattice constant $a_0$ ($\sim1$ \AA). This is exemplified in Fig.~\ref{fig:moire_hetero}, taken from Ref.~\cite{Zhang2017}. Panel A shows a topographic image of a WSe$_2$/MoS$_2$ heterobilayer acquired by atomic force microscopy, illustrating the layer stacking. Panel B displays a high-resolution probe image showing the moiré lattice with a periodicity close to 8.7 nm,
demonstrating the long-range periodic modulation induced by the small twist angle and natural lattice mismatch.

Although the resulting moiré pattern in a 2D heterostructure is largely determined by the crystal structure of its monolayer components and their relative alignment, it is also generally affected by sample-dependent factors. For instance, studies have shown that heterostrain can control moiré exciton minibands~\cite{Zheng2021twist}. Other factors, such as  lattice relaxation and defects, can lead to local distortions of the superlattice, or create domains of uniform stacking \cite{Yoo2019atomic,Lau2022reproducibility,Huang2022excitonsREVIEW,Zhao2023excitons}.

\subsection{Excitons in Moiré Potentials: Theory and Signatures}

In typical type-II semiconducting bilayers, the dominant term in the moir\'e potential for excitons comes from the spatial modulation of the electronic band gap \cite{Yu2017moire}. The band gap field $E_g(\vec{r})$ possesses minima where excitons can become localised, 
similarly to how optical lattices confine atoms in ultracold gases~\cite{Schafer2020}.
There are, however, two important departures from this picture. The first are hybrid excitons, for which the moir\'e potential can be dominated by the spatial modulation not of the band gap, but of the interlayer coupling that hybridises intra and interlayer excitons \cite{Alexeev2019resonantly,Shimazaki2020strongly,Polovnikov2024field}. Second, the Wannier-Mott paradigm can break down in moir\'e potentials where electrons and holes tend to localise at different positions \cite{Magorrian2021multifaceted}. If the depth of the moir\'e potential wells is comparable with the free-exciton binding energy, the electron and hole may still bind electrostatically while being spatially separated, forming so-called charge transfer excitons \cite{Naik2022intralayer}. The following discussion focuses on the simpler point-like Wannier-Mott excitons. Charge transfer excitons are further discussed in Sec.\ \ref{sec:ChTransf}.

In the case where the exciton Bohr radius is much smaller than the moiré superlattice period, the exciton can be regarded as a composite particle, with its centre of mass position $\vec{R}$ moving in an external periodic potential  $V_M(\vec{R})$ defined by the moiré potential. In type-II heterostructures, the intra and interlayer excitons are both well defined and $V_{M}(\vec{R})$ is simply a scalar potential. The low-energy exciton dynamics is governed by a single-particle Hamiltonian of the form~\cite{Wu2018theory}:
\begin{equation}\label{Eq.Wu2018}
    H_0 = \hbar\Omega_0 - \frac{\hbar^2 (\nabla_{\vec{R}}-i\vec{k}_0)^2}{2M} + V_M(\vec{R}),
\end{equation}
where $M$ is the total exciton mass, here assumed isotropic, as is the case in TMDs; $-i\hbar\nabla_{\vec{R}}$ is the centre-of-mass momentum operator, and $\hbar\Omega_0$ is the free exciton resonance energy, occurring for centre-of-mass momentum $\hbar\vec{k}_0$. For intralayer excitons, $\vec{k}_0=\mathbf{0}$, whereas for IXs it equals the valley mismatch, $\vec{k}_0=\vec{K}_{\rm top}-\vec{K}_{\rm bottom}\equiv \Delta \vec{K}$, between the electron and hole~\cite{AnomalousLightCones2015,Wu2018theory}. In homo-bilayers, as well as band-edge-matched heterostructures, interlayer tunnelling between electrons and holes can strongly mix intra- and interlayer excitons. This is modelled by the slightly more general Hamiltonian
\begin{equation}\label{eq:HforhXs}
H_0 = \left(
\begin{array}{cc}
\hbar \Omega_{\mathrm{X}} - \frac{\hbar^2 \nabla_{\vec{R}}^2}{2M_{\mathrm{X}}}
  + V_{M,\mathrm{X}}(\vec{R}) & T(\vec{R}) \\[4pt]
T^{*}(\vec{R}) & \hbar \Omega_{\mathrm{IX}} - \frac{\hbar^2 (\nabla_{\vec{R}}-i\Delta\vec{K})^2}{2M_{\mathrm{IX}}}
  + V_{M,\mathrm{IX}}(\vec{R})
\end{array}
\right).
\end{equation}
where the position-dependent tunnelling coefficient $T(\vec{R})$ mixes the two exciton species. Although moiré excitons are typically regarded as point-bosons, their internal structure, namely the electron and hole, can be directly time- and momentum- resolved ~\cite{Karni2022structure}.

The depth of the moir\'e potential determines whether the exciton spatially localises at specific sites of the moiré supercell, or whether it behaves as a delocalised exciton. Indeed, this transition has been studied theoretically ~\cite{Brem2023bosonic}, and observed experimentally by  using  $g$-factor measurements~\cite{Blundo2024localisation}. The ability to  image exciton confinement within a moiré unit cell with a subnanometer electron probe was demonstrated in Refs.~\cite{Susarla2022hyperspectral,Susarla2021mapping,Schmitt2022formation,Schmitt2025ultrafast}. Moiré confinement not only modifies the excitonic energy spectra, but also alters the exciton magnetic response. Ref.~\cite{Gobato2022distinctive} showed that moiré-confined intralayer excitons in MoSe$_2$/WS$_2$ heterostructures exhibit significantly reduced $g$-factors compared to free excitons, a consequence of their finite center-of-mass momentum and modified angular momentum composition in reciprocal space. This provides a magneto-optical fingerprint of exciton localisation within the moiré superlattice.

\begin{figure}[t]
\centering
\includegraphics[width=0.85\linewidth]{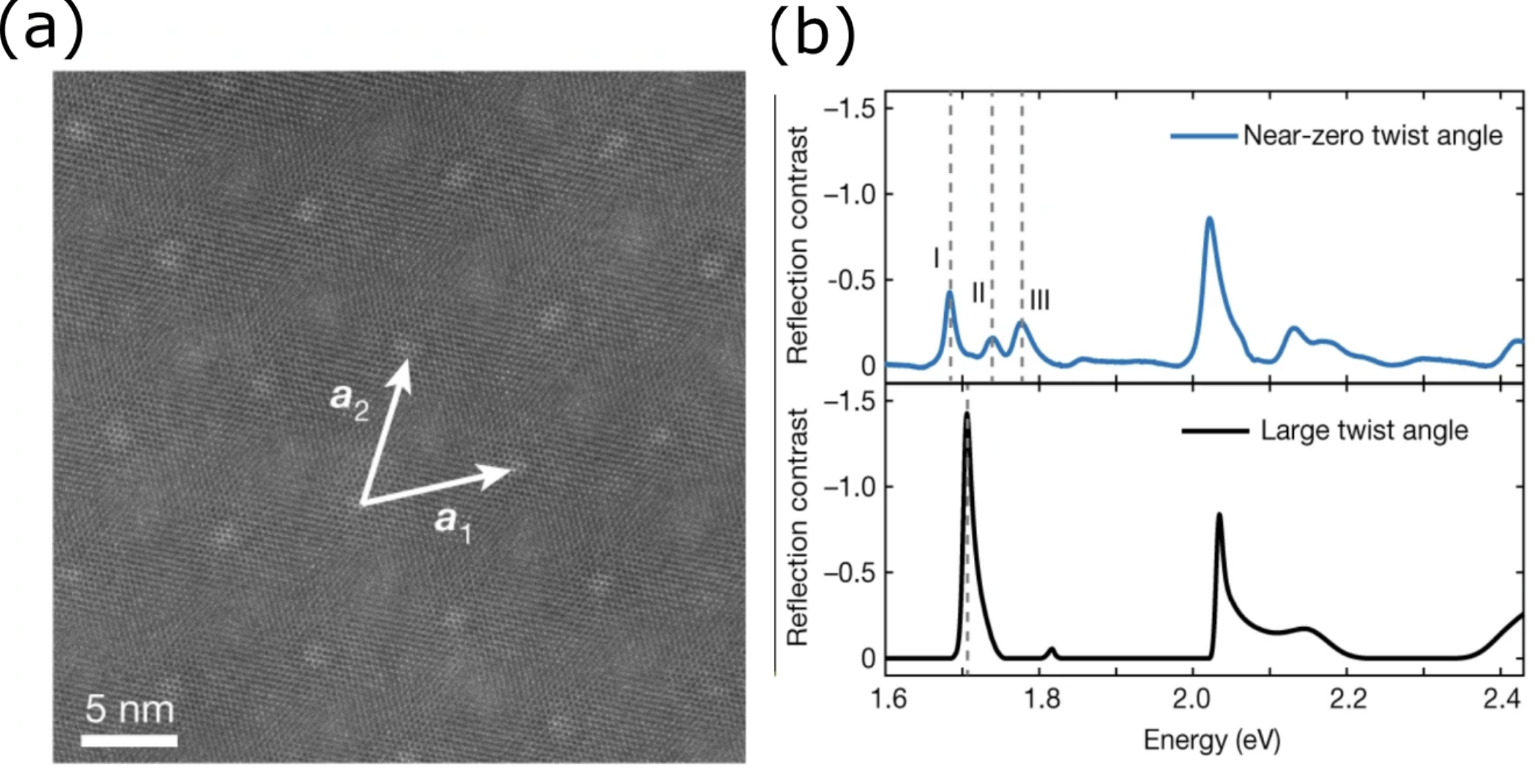}
\caption{\textbf{Moiré superlattice and excitonic fine structure in WSe$_2$/WS$_2$ heterobilayers.}
\textbf{(a)} Atomic-resolution scanning transmission electron microscopy (STEM) image of a near-zero twist angle WSe$_2$/WS$_2$ heterostructure, showing a well-defined moiré pattern with a periodicity of $\sim$8~nm. The lattice vectors $\vec{a}_1$ and $\vec{a}_2$ indicate the moiré superlattice orientation in real space.
\textbf{(b)} Reflection contrast spectra for WSe$_2$/WS$_2$ heterostructures with near-zero (top) and large (bottom) twist angles. For the near-zero twist configuration, the periodic moiré potential leads to the appearance of multiple excitonic resonances (labelled I–III) around the WSe$_2$. In contrast, the large twist angle sample displays only the primary intralayer exciton peak, confirming the absence of strong moiré coupling. These observations highlight the formation of flat excitonic bands due to moiré confinement in the strong-coupling regime. Figures adapted from \emph{Jin et al.} \cite{Jin2019observation}. 
}
\label{fig:moire_excitons}
\end{figure}

The behavior of moiré excitons can be understood from two complementary perspectives: momentum space and real space. In momentum space, the emphasis lies on the folding of exciton bands into moiré minibands within the mini-Brillouin zone of the moir\'e superlattice. In contrast, the real-space picture focuses on the formation of periodic exciton traps within the moiré lattice, leading to spatially localised exciton states with quantized energy levels. Proper understanding of both perspectives is necessary for capturing miniband formation, exciton hybridization, and the resulting optical selection rules.

\paragraph{Momentum-Space Picture.-} 
In reciprocal space, the moiré periodicity ``folds'' the free excitonic bands into a mini- or moir\'e-Brillouin zone (mBZ). The moir\'e potential mixes the folded bands by umklapp scattering, redistributing the exciton's oscillator strength---formerly finite only for exciton momenta near zero---amongst multiple exciton minibands at the mBZ centre. This introduces new optically active states, visible as additional peaks in photoluminescence and reflectance contrast spectra~\cite{Jin2019observation, Alexeev2019resonantly,Seyler2019signatures}. 
The resulting multi-peak optical spectra is currently regarded as a smoking gun for miniband formation, and as direct evidence of moiré-modified excitonic band structures~\cite{Huang2022excitonsREVIEW, Jin2019observation}.

 Figure~\ref{fig:moire_excitons}, taken from Ref.~\cite{Jin2019observation}, illustrates these phenomena in WSe$_2$/WS$_2$ heterostructures. Panel (a) shows atomic-resolution STEM images of a moiré superlattice with $\sim$8~nm periodicity, consistent with a near-zero twist angle. In the optical domain, in Fig.~\ref{fig:moire_excitons}(b), the reflectance contrast spectrum displays a dramatic difference between small and large twist angles. At large twist angles, the spectrum only exhibits a single exciton peak corresponding to an essentially unperturbed WSe${}_2$ intralayer $A$ exciton. At small twist angles, a moiré pattern forms,  and multiple pronounced resonances appear in that same energy region. These peaks are signatures of the moiré minibands formed by strong confinement of the intralayer excitons by potential wells in the moiré landscape, which only happens at small twist angles, when the potential wells are wide enough to confine the excitons. The folding to the mini-BZ is further explained in Fig.~\ref{figmoirebands}, where the spatially periodic hybridisation between intra- and inter-layer excitons leads to the formation of a mBZ, of flatbands, and of additional bright states, which we will detail in Sec.~\ref{twistronics}.

\paragraph{Real-Space Picture.-}
In real space, one may focus on the moiré potential minima appearing periodically across the sample, each constituting an exciton trap \cite{Wu2014}. These traps can be as deep as $100-200~meV$, with sizes on the order of 10~nm~\cite{Tran2019evidence, Seyler2019signatures,Lin2023remarkably,Zheng2023strong} and point symmetry inherited from the moir\'e potential \cite{Yu2017moire,Brotons2020spin}. 
The appearance of such periodic traps has been visualized,  as shown in Fig.~\ref{fig:moire_excitons}(a), where they appear as bright spots appearing across the sample with the periodicity of the moiré superlattice. It is worthwhile mentioning that the moir\'e potential minima tend to appear at regions of the moir\'e supercell with highly symmetric atomic registries \cite{Yu2017moire}.

Exciton confinement leads to a discrete, quantum-dot-like spectrum of energy levels, corresponding to the multiple confined exciton states, with wave functions that behave as irreducible representations of the potential well's point symmetry group \cite{Ruiz2020theory}.
 The discrete energy spectrum of confined excitons explains the multi-peak optical spectra already  described~\cite{Tran2019evidence,Seyler2019signatures,Jin2019observation,Alexeev2019resonantly}. Moreover, the symmetry of the localised wave functions, combined with that of the underlying Bloch states, results in specific optical selection rules that may differ substantially from those of the monolayers\cite{Seyler2019signatures,Ruiz2020theory}, as discussed below.

Finally, as the twist angle increases, the moiré potential wells shrink in size, and eventually also shallow, leading to the delocalisation of excitons across the sample. Moreover, the discrete energy levels broaden and connect into dispersive minibands~\cite{Magorrian2021multifaceted,Stansbury2021visualizing,Zheng2021twist}, thus bridging the real-space and momentum-space pictures.

\paragraph{Symmetry and Optical Selection Rules.} The moiré potential not only modifies the exciton energy landscape, but also imprints spatial variations into their optical selection rules. Once the exciton localises at a given site of the moir\'e supercell, the local atomic stacking determines the symmetry of the electron and hole Bloch functions, and the shape of the potential well determines the symmetry of the envelope function. Thus, each potential localisation site has its own selection rules, leading to a spatially varying optical dipole orientation across the moiré pattern \cite{Yu2017moire}. In TMD heterobilayers, moir\'e excitons tend to localise at regions with high-symmetry local stacking. The $C_3$ symmetry of the underlying atomic lattices is preserved about these points, and the trapped exciton wave functions become eigenstates of this rotation, with eigenvalues that directly relate to the local optical selection rule, i.e., to whether excitons localised at that region interact with light of left ($\sigma^-$) or right ($\sigma^+$) circular polarisation, or even out-of-plane linear polarisation ($\sigma^z$)~
\cite{Jin2019identification,Yu2017moire, Seyler2019signatures,Ruiz2020theory,Ge2023observation}. 
Although the actual localisation site is fully determined by the moir\'e potential minima, the latter can be modified by electrical means~\cite{Yu2017moire, Wilson2021}, leading to unprecedented electric control over the optical response of exciton states.

\subsection{Optical spectra of moir\'e excitons: quantum emitter arrays}

Moir\'e excitons were first identified in TMD heterobilayers through their multi-peak structures in photoluminescence and reflectance contrast spectra. These optical features are common to intralayer \cite{Jin2019observation}, interlayer~\cite{Seyler2019signatures,Tran2019evidence} and hybridised excitons \cite{Alexeev2024nature}.  Nonetheless, these excitonic species exhibit distinct dependencies on external parameters such as twist angle~\cite{Alexeev2019resonantly,Seyler2019signatures,Tran2019evidence,hidalgo2023interlayer,roman2023excitons}, magnetic fields~\cite{Seyler2019signatures,She2025magneto}, and carrier density~\cite{Jin2019observation}, and display markedly different transport behaviours~\cite{Knorr2022,Fowler2021voltage,Wietek2024,Shentsev2025,Qian2023strongly,Conti2023,Liu2023direct,Rossi2024anomalous,Ray2025diffusion,Choi2020moire}, as well as selection rules for optical transitions~\cite{Wu2024revealing,Xie2024emergence,Michl2022intrinsic}.

Intralayer moir\'e excitons were observed in WSe$_2$/WS$_2$ heterostructures~\cite{Jin2019observation}, where the absorption spectra exhibited three peaks around the intralayer A exciton resonance of WSe$_2$.  Two of these exciton states  show pronounced blueshifts and reduced oscillator strength upon electron doping, in contrast to the weak response expected from conventional screening. This anomalous behavior indicates that these states are spatially localised at moiré potential minima—real-space regions that also serve as preferential sites for gate-induced electrons.

Interlayer moir\'e excitons were first observed in MoSe$_2$/WSe$_2$ heterostructures~\cite{Seyler2019signatures,Tran2019evidence}. Looking at samples with $\theta\approx1^\circ$ and $\theta\approx2^\circ$, their peak energy $E_M$ was shown to slightly increase with $\theta$, which was interpreted as a signature of the exciton localisation being twist-dependent~\cite{Tran2019evidence}. Because the size of the moir\'e unit cell goes approximately as $l\propto \theta^{-1}$, the increase in $E_M$ at larger angles is consistent with narrowing potential wells, leading to weaker lateral confinement. The magnetic properties of interlayer moir\'e excitons were also probed by the application of magnetic fields~\cite{Seyler2019signatures}, showing that the valley alignment determined by $\theta$ leads to the $g$ factor taking either of two values, attributed to the exciton transitions occurring within the same or different valleys.

\begin{figure}[t]
    \centering
    \includegraphics[width=0.95\textwidth]{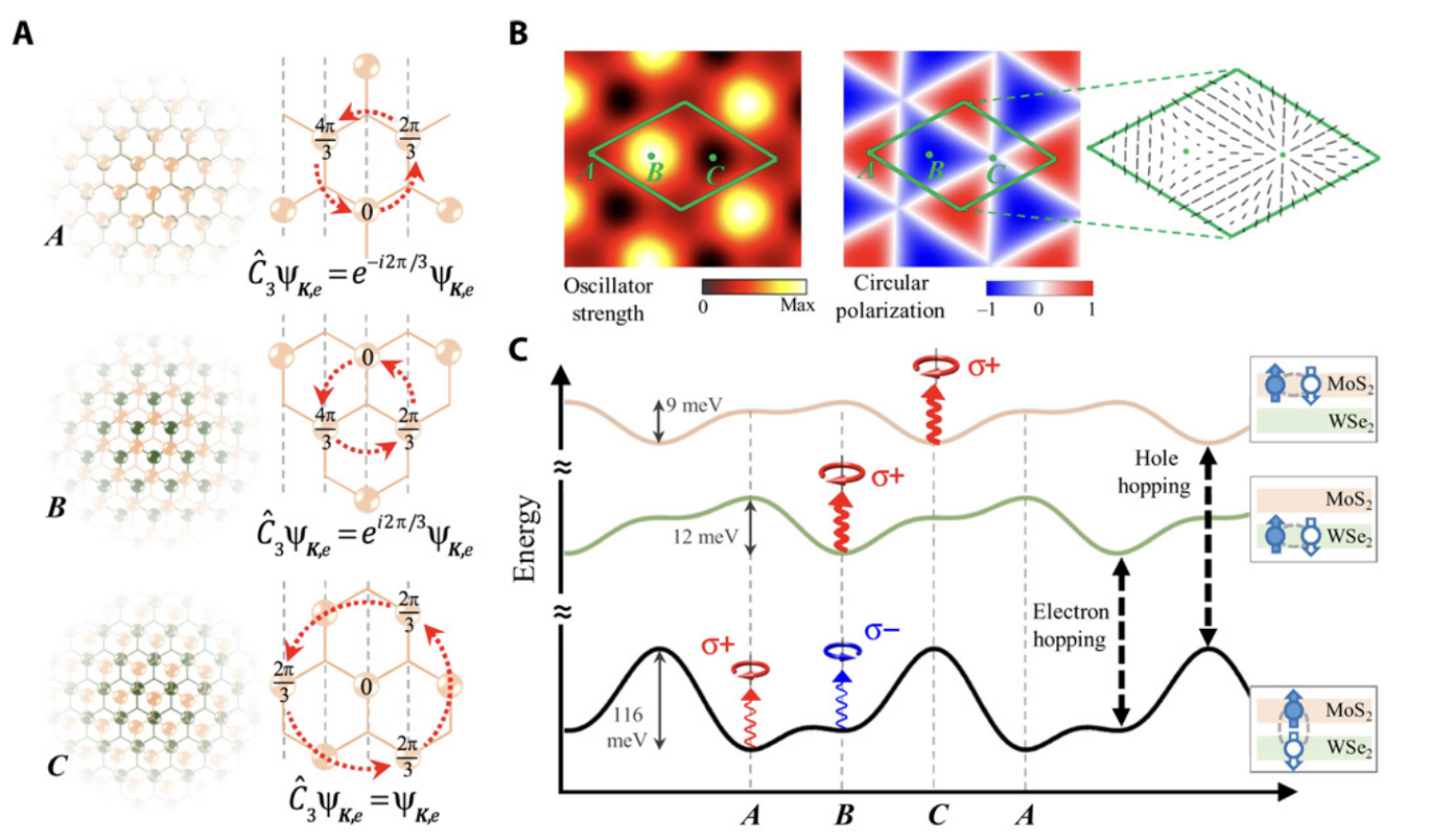}
    \caption{\textbf{Nanopatterned spin-optical properties of interlayer moiré excitons.} (\textbf{A}) Symmetry properties of exciton wavefunctions at three high-symmetry registries (A, B, C), showing distinct $C_3$ rotation eigenvalues. (\textbf{B}) Left: Spatial modulation of oscillator strength. Center: Circular polarization map of exciton emission, with opposite helicities at A and B sites. Right: Ellipticity of intermediate sites. (\textbf{C}) Schematic exciton potential landscape, showing localised exciton energy minima and helicity-dependent transitions. The energy difference between A and B sites is tunable via an external electric field, allowing for programmable quantum emitter arrays. Figures adapted from \emph{Yu et al.}~\cite{Yu2017moire}. }
    \label{fig:nanopatterned_spin_optics}
\end{figure}

    As shown in Fig.~\ref{fig:nanopatterned_spin_optics}, different local stacking registries give rise to position-dependent optical selection rules and oscillator strengths, leading to a nanoscale pattern of circularly polarized quantum emitters \cite{Yu2017moire,Guo2021moire}. These emitters are associated with energy minima in the moir\'e potential landscape, where interlayer excitons are confined in discrete traps and exhibit site-specific optical helicities. The periodic energy landscape can support not only localised emission but also quantum tunneling between minima, giving rise to miniband formation or hybridisation with intralayer excitons \cite{Zheng2025Forster}. 

Hybridised moir\'e excitons were first observed in MoSe$_2$/WS$_2$ heterostructures~\cite{Alexeev2019resonantly}, and more recently in MoTe${}_2$/MoSe${}_2$ heterostructures \cite{hXMoTe22024}, both based on theoretical predictions of strong hybridisation between near-resonant intra and interlayer excitons~\cite{Ruiz2019interlayer}. Analysing multiple samples with different twist angles, a prominent exciton red shift was observed close to lattice alignment ($\theta\sim 0^\circ$) and anti-alignment ($\theta\sim60^\circ$), an indicator of exciton moir\'e localisation. However, in this case, the authors argued, localisation was driven not by a modulation of the heterostructure band gap, but of the position-dependent hybridisation between intra and interlayer excitons. In MoSe${}_2$/WS${}_2$ (MoTe${}_2$/MoSe${}_2$), the lowest energy IX is close to resonance with the intralayer A exciton of the MoSe${}_2$ (MoTe${}_2$) layer, thus promoting hybridisation between the two species as described by the Hamiltonian \eref{eq:HforhXs}, in these cases mediated by interlayer electron tunnelling. 
Hybridisation gives rise to low-energy anti-bonding states, which constitute potential wells for excitons. and are deepest at sample regions where the metal atoms of both layers align vertically in both aligned and anti-aligned structures.  A natural mechanism to tune the hybrisidation of moiré excitons is the quantum-confined Stark effect \cite{Tang2021tuning,Hagel2022electrical}. This affects IXs exclusively, allowing to control their detuning with intralayer excitons, and thus the degree of hybridisation. Further evidence for the hybridisation mechanism of moir\'e exciton formation was recently obtained by this method \cite{Shimazaki2020strongly,Polovnikov2024field}.

\subsection{Twistronics with moiré excitons}
\label{twistronics}
The twist angle between layers in a van der Waals heterostructure is a powerful tuning knob for electronic and excitonic band structures, a concept broadly referred to as \textit{twistronics}~\cite{Carr2017,Hennighausen_2021,Wang2025twist,Rosenberger2020twist,Hsu2019tailoring,Orazbay2024,Liao2020precise}. In 2D semiconductor bilayers, the twist angle modifies not only the moiré superlattice periodicity, but this is accompanied by a change in the exciton hybridisation \cite{Alexeev2019resonantly,McDonnell_2021,Tang2021tuning,Sokolowski2023twist},  recombination dynamics \cite{Choi2021twist,Chen2022controlling,Kim2023dynamics,Cai2023interlayer,Roux2025exciton}, diffusion \cite{Zhong2025twist,Ray2025diffusion}, optoelectronic response
~\cite{Villafañe2023twist,Zhang2020twist, Seyler2019signatures,Zhao2021dynamic, Zheng2021twist,Zheng2022thickness,Zheng2023evidence,Woo2023excitonic,Liu2024enhanced,Palekar2024anomalous,Wang2025twist,Dhakal2025giant,Zeng2025small,Tian2025regulating}, and even the effective dimensionality \cite{Kennes2020one,Wang2022one,Soltero2023dimensionality,Zhao2023excitons,Guo2023pseudo}.

In Ref.~\cite{Alexeev2019resonantly}, varying the twist angle leads to a modification of the hybridisation between intra and inter-layer excitons. This hybridization is strongest near twist angles of $\theta \approx 0^\circ$ and $60^\circ$, where the Brillouin zones align or anti-align, and zone-centre excitons of the two species are nearly resonant. 
Hybridised exciton states display energy shifts up to tens of meV with twist angle, and exhibit enhanced or suppressed oscillator strengths, depending on their intralayer exciton content \cite{Ruiz2019interlayer}. 

\begin{figure}[t]
\centering
\includegraphics[width=0.95\linewidth]{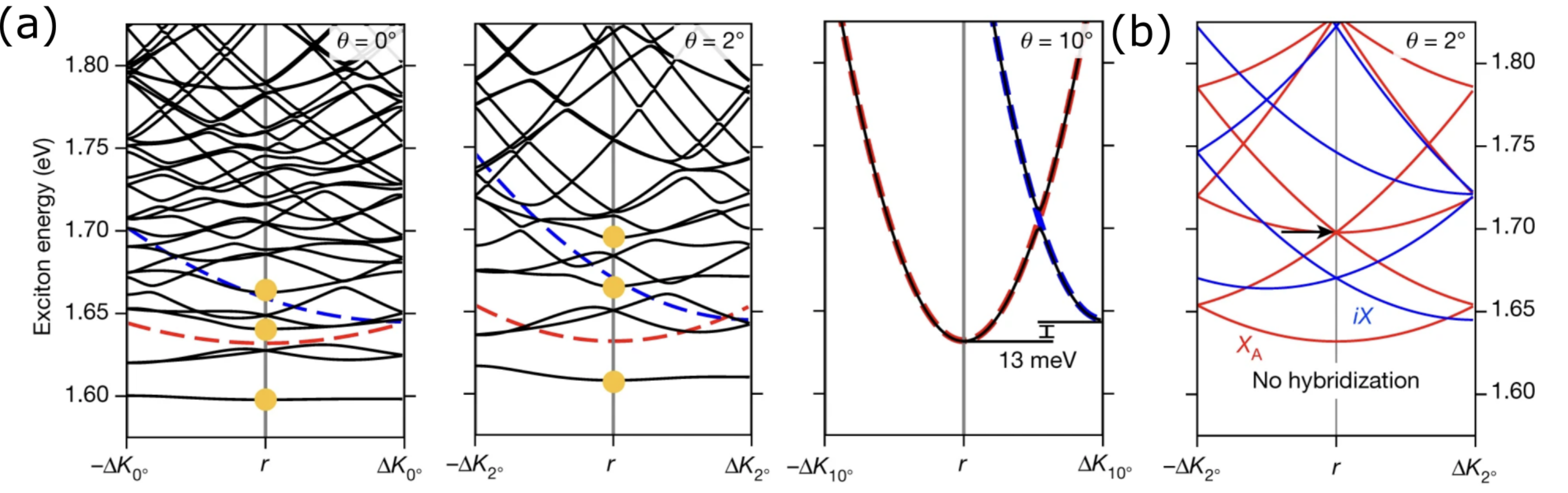}
\caption{Twist-angle-dependent evolution of exciton band structures in MoSe$_2$/WS$_2$ heterobilayers. (a) Moiré minibands of hybridized excitons calculated for $\theta = 0^\circ$, $2^\circ$, and $10^\circ$. At small angles, avoided crossings indicate strong hybridization between intralayer (red dashed) and interlayer (blue dashed) excitons. (b) Band structure in the absence of tunneling, showing no hybridization. Figures adapted from \emph{Alexeev et al.}~\cite{Alexeev2019resonantly}.}
\label{figmoirebands}
\end{figure}

Figure~\ref{figmoirebands} shows the evolution of the exciton band structure as a function of the twist angle reported in Ref.~\cite{Alexeev2019resonantly}, for  a MoSe$_2$/WS$_2$ heterobilayer. At small angles, strong hybridisation leads to moiré minibands with flat dispersions and large energy gaps, corresponding to well localised exciton states. With increasing twist angle, hybridisation weakens and the exciton bands become more dispersive, and essentially decoupled.  In the absence of the hybridisation mechanism (panel b), intralayer and interlayer excitons remain energetically separate and uncoupled, underscoring the role of interlayer coupling in driving moiré physics.

In addition to the hybridisation of the exciton branches and the formation of the mBZ, the twist angle can also modify the Landé $g$-factor of interlayer excitons, leading to a change in valley composition and orbital character~\cite{Seyler2019signatures}. These excitons also exhibit strong circular dichroism and valley polarization near high-symmetry angles, governed by the stacking-induced symmetry breaking. The exciton lifetimes also can vary by even an order of magnitude with the twist angle~\cite{Choi2021twist}, as the degree of momentum and spin alignment across layers controls the radiative recombination pathways.

Together, these observations establish the twist angle as a critical parameter for engineering moiré excitons with tailored optical properties and miniband structures, forming the foundation for future twistronic control in bosonic quantum matter.

\subsection{Lattice reconstruction and its effects on the moir\'e potential}
Although the moir\'e patterns found in 2D heterostructures are often imagined as ideal, i.e., arising from interference between rigid lattices, real samples can exhibit important amounts of reconstruction, especially for marginally small twist angles and lattice mismatches. Lattice reconstruction occurs when the adhesion energy gained from creating domains with a specific interlayer atomic registry overcomes the elastic energy cost of the associated lattice deformation \cite{Yoo2019atomic,Lau2022reproducibility,Huang2022excitonsREVIEW,Li2023lattice,EnaldievPRL,Zhao2023excitons}. For TMD structures, this situation naturally arises in well aligned (twist angles $\lesssim 2^\circ$) homobilayers \cite{Weston2020}, or chalcogen-matched heterobilayers\cite{Weston2020,Rosenberger2020}, which have nearly identical lattice constants, thus producing large-periodicity moir\'e patterns where strain can be distributed throughout a larger surface area. Lattice reconstruction fundamentally alters the description of the electronic and excitonic properties at small twist angles, from one where regions of different atomic registry are evenly distributed throughout the sample, to another of alternating low energy configurations, separated by narrow domain walls where rich one-dimensional \cite{Soltero2024,MoulsdaleDomains} and zero-dimensional \cite{Soltero2024} physics has been predicted. 
Relaxation can still play an important role in moir\'e materials at larger twist angles, as observed in recent studies on MoTe$_2$ twisted bilayers probing the fractional quantum anomalous Hall effect at $\theta\sim4^\circ$ \cite{Cai2023signatures,Xu2024Observation}.


\subsection{Beyond the Wannier-Mott paradigm: charge-transfer moir\'e excitons}\label{sec:ChTransf}
Continuum models have played a central role for understanding the electronic and excitonic properties of moiré heterostructures. These effective models describe the low-energy bands of carriers or excitons using smooth potentials derived from symmetry and interlayer coupling, circumventing the complexity of full atomistic calculations. These models have successfully captured moiré-induced minibands, hybridisation effects, and selection rules across a wide range of materials and twist angles~\cite{Bistritzer2011moire,HongyiPRB2017,Yu2017moire,Wu2017topological,Wu2018theory,Ruiz2019interlayer,FengchengWuPRL2019}. Theoretically, it is known that the moir\'e potentials for electrons and holes are, in general, different, exhibiting potential wells at different sites of the moir\'e supercell \cite{Magorrian2021multifaceted,Soltero2022moire,Hagel2024polarization}, separated by nanometric distances. In spite of this, a long standing assumption has been that, even in moir\'e excitons, the electron and hole would remain close together, given the  large exciton binding energies found in 2D semiconductors. An \emph{ansatz} emerged from this picture, whereby the moir\'e exciton Bohr radius is assumed much smaller than the moir\'e periodicity. The moir\'e excitons are then described by the Wannier-Mott equation \eref{eq:wanniermott}, supplemented by an exciton moir\'e potential
\begin{equation}
    V_X(\rr_e,\,\rr_h)\approx V_e(\RR)+V_h(\RR),
\end{equation}
where both the electron and hole positions are taken at the COM coordinates $\RR$ \cite{HenriquesXPhos,Ruiz2020theory,Soltero2023dimensionality}.  
However, recent \emph{ab initio} calculations on aligned WSe$_2$/WS$_2$ heterostructures have shown that an alternative picture~\cite{Naik2022intralayer} is possible, where electrons and holes localised at different moir\'e sites can bind electrostatically into so-called \emph{charge transfer moir\'e excitons}. By now, experimental evidence exists for both intralayer~\cite{Naik2022intralayer,Li2024imaging} and interlayer~\cite{Wang2023intercell} charge transfer moir\'e excitons in the TMD heterobilayer WSe${}_2$/WS${}_2$.

The existence of intralayer charge-transfer excitons was confirmed experimentally by their distinct optical signatures. For example, whereas modulated Wannier-Mott moiré excitons exhibit enhanced oscillator strength and tight electron-hole correlation near high-symmetry stacking regions, charge transfer excitons exhibit reduced oscillator strength and strong sensitivity to environmental screening due to their extended spatial character. This can be observed experimentally via reflection spectroscopy \cite{Naik2022intralayer}, while their spatial distribution can be mapped by photocurrent tunneling microscopy \cite{Li2024imaging}.

Interlayer excitons interact strongly, both amongst themselves and with free charge carriers~\cite{Li2020dipolar,Yu2021,Brotons2021moire}, due to their permanent out of plane electric dipole moment. In addition to this, interlayer charge-transfer excitons possess permanent quadrupole moments due to the in-plane separation between the electron and hole, thus enhancing their interaction with charge carriers. Ref.~\cite{Wang2023intercell} demonstrated twistronic control over this interaction by contrasting the optical responses of interlayer excitons in aligned and anti-aligned, hole-doped WSe${}_2$/WS${}_2$ heterostructures. For the holes, the combined moir\'e potential and Coulomb repulsion led to the formation of generalised Wigner crystals: periodic arrays of localised holes, with a lower crystal symmetry than that of the moir\'e pattern. Photoexcited moir\'e interlayer excitons would then interact with the crystal, and either bind with it, if their net interaction was sufficiently attractive, or remain free if the interaction was weak. The former case was realised for interlayer charge transfer excitons due to their large in-plane quadrupole moment, exhibiting a red shift of $\approx\,7\,{\rm meV}$ as it bound to the Wigner crystal. Crucially, that type of IX only appears in the anti-aligned case, where the electron and hole moir\'e sites do not coincide~\cite{Wang2023intercell}.

Moir\'e excitons have been observed in a variety of settings beyond what we have already discussed, including homo-bilayers, homo- and heterostructures with more than two layers, as well as in 2D semiconductors other than TMDs~\cite{Forg2021moire,Andersen2021excitons,Zhao2021strong,Wu2022evidence,Wu2022observation,Li2022giant,Lian2023exciton,Zheng2023localization,Zheng2023exploring,Wu2023effect}, highlighting the ubiquity and profound importance of moir\'e excitons in the physical properties of van der Waals materials.
 

\subsection{Rydberg excitons}
Rydberg moiré excitons arise from the interplay between highly excited excitonic states and the periodic moir\'e potential~\cite{Hu2023observation}. Due to their large Bohr radius and enhanced polarizability, Rydberg excitons are particularly sensitive to the moiré superlattice. Recent experiments have demonstrated that Rydberg excitons in monolayer WSe$_2$ can be spatially trapped by the Coulomb landscape generated by an adjacent twisted bilayer graphene (TBG). This {\it indirect} process effectively induces a moiré Rydberg exciton ~\cite{Hu2023observation}. In the strong-coupling regime where the moiré wavelength exceeds the exciton radius, spectroscopic features such as energy splitting, linewidth narrowing, and redshift appear, highlighting their hybrid character and charge-transfer nature.

Recent experiments have resolved the excitonic energy spectrum and mapped out the full three-dimensional spatial profiles of Rydberg excitons—including their in-plane periodicity and out-of-plane nodal structures—demonstrating full wavefunction tomography in a moiré-engineered potential landscape~\cite{Zhao2024atomic}. Further experiments in twisted bilayer WSe$_2$ revealed signatures of Rydberg excitons~\cite{He2024dynamically}.

Moiré Rydberg excitons open new paths for exploring quantum many-body phenomena, nonlinear optics, and topological excitonic states in engineered moiré quantum materials. Owing to their large spatial extent and strong dipole moments, they interact strongly with the periodic moiré potential and with other excitons, enabling enhanced nonlinearities, long-range interactions, and the formation of strongly correlated excitonic states within moiré minibands.

\subsection{Moir\'e Phonons and Excitons in Twisted Heterostructures.} 

The formation of moiré superlattices in twisted TMDs not only redefines the electronic and excitonic bands, but can also dramatically modify the lattice dynamics~\cite{Guo2023pseudo,Du2019,Lin2018moire,Parzefall2021moire,Shinokita2021resonant,Chuang2022,Rahman2022,Lim2023modulation,Jürgens2024,Li2025moire,Santos2025enhanced}. The periodic modulation of the atomic registry gives rise to moiré-folded phonons that originate from the zone folding of the phonon dispersion into the mini Brillouin zone. These moiré-induced lattice vibrations have been experimentally observed~\cite{Shinokita2021resonant,Rahman2022,Pimenta2023pressure} to exhibit both frequency shifts and Raman intensity enhancement as a function of twist angle. Importantly, moiré phonons can couple selectively to excitonic transitions, providing a unique spectroscopic fingerprint of the moiré potential and its symmetry.  This coupling can manifest in resonant Raman scattering processes or phonon-assisted exciton relaxation, offering new pathways to probe exciton localisation and inter-well dynamics within the moiré landscape. Chiral phonons have also been demonstrated to flip the angular momentum of excitons~\cite{Delhomme2020flipping}. Moreover, the phononic properties of moiré materials can be actively tuned through external parameters such as pressure, strain, and electrostatic gating. Combined with valleytronics and twistronics, these control knobs allow for dynamic modulation of moiré phonon spectra and their interactions with excitons, enabling new regimes of exciton-phonon physics.

Moiré excitons are not exclusive to TMD heterobilayers. Indeed, under hydrostatic pressure, the twisted bilayer graphene also supports tightly bound excitons as shown in Ref.~\cite{Duarte2024moire}. Solving the Bethe-Salpeter equation on top of GW-corrected electronic bands, they show that their quasiparticle properties, namely the binding energy, spatial extend, and dipole moment, can be controlled externally.

\section{Quantum Many-Body Phases with Moiré Excitons}

\subsection{Moiré excitons as realizations of Bose-Hubbard models}

It was early recognized that moiré superlattices formed by stacked TMD layers could serve as quantum simulators of Hubbard-like systems. This arises from the moiré-induced localisation of charge carriers and excitons to the emergent superlattice, which accentuates the Coulomb interactions by suppressing kinetic energy. The ability to tune both the periodic potential and interaction strength through external parameters, such as twist angle, gating, pressure, and dielectric environment opens new possibilities to engineer diverse classes of Hubbard models in which strongly correlated phases of matter can arise. This unprecedented control has established moiré heterostructures as one of the most promising solid-state platforms for quantum simulation~\cite{Wu2018Hubbard,Tang2020simulation,Kennes2021moire,Xu2022A}.

Fermi-Hubbard models are naturally realized with electrons and holes in these systems, and the moiré pattern allows the realization of various lattice geometries \cite{Finney2019tunable,Kennes2021moire,Angeli2021gamma}, including honeycomb \cite{Magorrian2021multifaceted,Campbell2024the}, triangular \cite{Tang2020simulation,Wang2020correlated}, rectangular \cite{Kennes2020one,Wang2022one,Soltero2022moire}, and Kagome lattices~\cite{Angeli2021gamma}. These tunable geometries have enabled the experimental observation of correlated electronic phases such as superconductivity, density waves~\cite{Slagle2020,Jin2021stripes}, Mott insulators \cite{Tang2020simulation}, and Wigner crystals~\cite{Regan2020mott}, and have supported the theoretical prediction of exotic quantum states like spin liquids, Majorana fermions, and topological or magnetically ordered phases. The study of strongly correlated phases in moiré systems remains a vibrant and growing research area~\cite{Wu2018Hubbard,Vitale2021,Lagoin2021key,Kennes2021moire,Du2023moireREVIEW,Chen2022tuning,Wang2022light}.

Moreover, moiré heterostructures provide a versatile platform to go beyond the standard Hubbard framework. Recent theoretical and experimental developments have shown the feasibility of simulating multi-orbital lattice models \cite{Campbell2024the}, asymmetric $p_x$–$p_y$ orbital systems, and one-dimensional–two-dimensional crossover scenarios in rectangular or low-symmetry geometries \cite{Kennes2020one,Wang2022one,Soltero2022moire,Soltero2023dimensionality,Guo2023pseudo}. These advances offer unprecedented opportunities to explore quantum criticality, unconventional pairing mechanisms such as chiral $d+id$ superconductivity and topological transitions within a highly controllable condensed-matter architecture.

On the other hand, due to their bosonic nature, excitons manifest distinct collective phenomena in the presence of a moiré superlattice. In particular, interlayer excitons, comprising an electron and a hole confined in different layers, possess a permanent out-of-plane dipole moment. When combined with spatial localisation induced by the moiré potential, this gives rise to extended Bose–Hubbard models featuring tunable lattice confinement and strong dipole–dipole interactions~\cite{Brem2020tunable,Li2020dipolar,Götting2022,Fang2023localization,Chatterjee2023harmonic,Lin2023remarkably}, providing a highly controllable platform to explore exotic excitonic phases \cite{Zeng2022strong,Julku2024exciton,Kiper2025confined,Huang2025collective}, including superfluids~\cite{Lagoin2021quasicondensation,Remez2022leaky,Deng2022moire,Zhang2022doping,Kwan2022excitonic,Santiago2023collective}, supersolids~\cite{Julku2022nonlocal,Camacho2022optical}, Mott insulators~\cite{Zhang2022doping}, density waves \cite{Zeng2023exciton}, dipole ladders \cite{Park2023dipole}, and different types of excitonic insulators~\cite{Zhang2022correlated,Gu2022dipolar,Chen2022excitonic,Xu2022A,Xiong2023correlated,Gao2024excitonic,Lian2024valley,Tan2023layer}, as well as magnetic ~\cite{Yang2024exciton} and topological phases~\cite{Stefanidis2020exciton,Xie2024long,Froese2025}. Moreover, these systems also exhibit rich dynamical behavior~\cite{Liu2020direct,Lavor2021zitterbewegung,Chen2024spatial,Deng2025frozen,Arsenault2025time}, and their strong exciton-exciton interactions can be used to control the valley Zeeman effect of the carriers~\cite{Li2021optical}.

The strong confinement of excitons to moiré sites has enabled the observation of single-photon quantum emitters~\cite{Shanks2021,Yu2017moire}, cascade transitions~\cite{Tan2022signature}, and  collective light–matter states~\cite{Baek2020highly,Qian2024lasing,Kumlin2025}. Interestingly, in moiré lattices, excitons can exhibit departures from ideal bosonic statistics~\cite{Lohof2023confined,Huang2023mott,Huang2024}.

Recent experiments on \textit{H}-stacked WS$_2$/WSe$_2$ heterobilayers have demonstrated that interlayer moiré excitons can form many-body bound states with surrounding charge lattices, leading to intercell moiré exciton complexes with distinctive spectral shifts and polarization signatures~\cite{Wang2023intercell,Chen2023excitonicREVIEW}.

Although the exciton-exciton interaction in van der Waals heterobilayers is usually modelled as a dipole-dipole repulsion, recent studies suggest that the fermionic substructure of excitons can lead to important corrections \cite{Katzer2023exciton,Steinhoff2024exciton,Huang2024nonbosonic}. Renormalization effects introduced by exchange, Pauli-blocking and screening can be strong enough to result in an effective attractive interaction at low densities \cite{Steinhoff2024exciton}. Therefore, there are regimes where exciton-exciton interactions can go beyond the usual picture of dipole repulsion.

\subsection{Experimental Realization of Excitonic Insulators}

Recent experiments have demonstrated the realization of excitonic insulators in moiré heterostructures~\cite{Gu2022dipolar,Zhang2022correlated,Chen2022excitonic,Xiong2023correlated,Gao2024excitonic,Lian2024valley,Mhenni2024gateArxiv}. These systems exploit the long lifetimes and permanent dipole moments of interlayer excitons to probe strongly correlated bosonic phases. A key strategy in these experiments involves doping moiré bilayers to a filling factor of one, thereby inducing a Mott insulating state. Upon applying a vertical electric field, carriers are redistributed: electrons remain in the bilayer, while holes are transferred to a neighboring monolayer, enabling the formation of tightly bound interlayer excitons stabilized by strong Coulomb attraction.

\begin{figure}[b]
\includegraphics[width=.75\columnwidth]{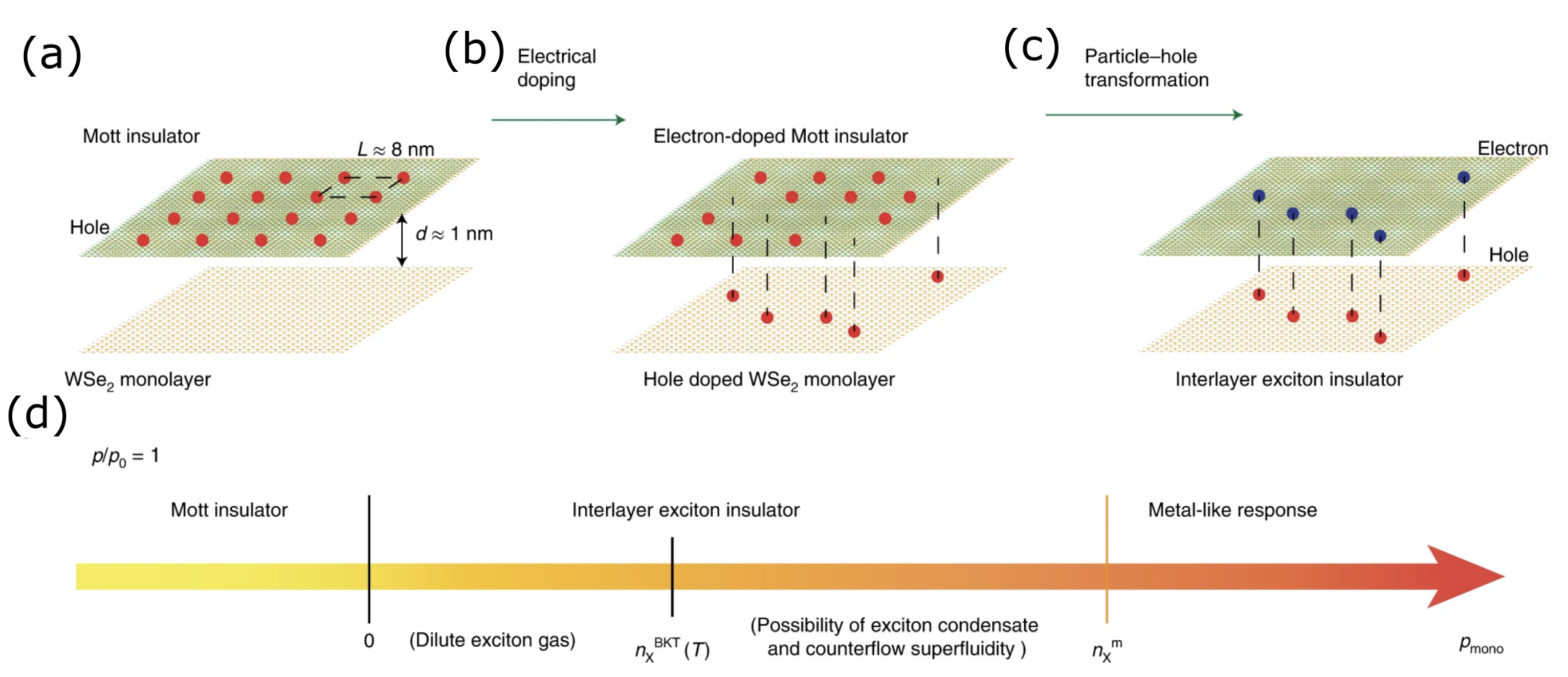}
\caption{Correlated interlayer exciton insulator and associated phase diagram in moiré double layers.
 Schematic evolution of electronic configurations in a WSe$_2$/WS$_2$/WSe$_2$ heterostructure as a function of electrical doping and particle-hole transformation. 
(a) The initial state corresponds to a Mott insulator in the WS$_2$/WSe$_2$ moiré bilayer, where each site hosts a single hole. 
(b) Upon electrical doping, additional holes populate the adjacent WSe$_2$ monolayer, leading to a partially doped Mott insulator. 
(c) A particle-hole transformation reveals this system as an interlayer exciton insulator formed by Coulomb-bound pairs between electrons in the doped moiré bilayer and holes in the monolayer. 
(d) Phase diagram of the correlated double-layer system at fixed total filling, illustrating the evolution from a pure Mott insulator to an interlayer exciton insulator and eventually to a metallic regime as the hole density in the monolayer $p_\mathrm{mono}$ increases. Intermediate exciton densities offer the possibility of interlayer exciton condensation and counterflow superfluidity.
 (Figure adapted from Zhang \textit{et al.}~\cite{Zhang2022correlated}).}
\label{fig:Zhang2022}
\end{figure}

A dipolar excitonic insulator was realized in a {\it double-layer} configuration, where a WSe${}_2$ monolayer is coupled via Coulomb interaction to a  WSe${}_2$/WS${}_2$ moiré bilayer~\cite{Gu2022dipolar,Zhang2022correlated}. The bilayer is a Mott insulator at one hole per moir\'e site. Upon the presence of an external field, holes are transferred to the WSe${}_2$ monolayer. This produces electrons in the moiré bilayer and holes in the monolayer that remain strongly bounded, forming a dipolar excitonic insulator. Although the state remains charge insulating, excitons can hop around in the lattice. As the density of holes in the monolayer is increased, the excitonic insulator persists until reaching a critical hole concentration where excitons dissociate. 
The phase diagram shows a crossover from a Mott insulator to a correlated exciton fluid, and ultimately to a compressible metallic state. The real-space configuration and phase-diagram are illustrated in Fig.~\ref{fig:Zhang2022}. 
The phase-diagram shows how the system evolves from a Mott insulator to a metallic phase through an excitonic insulator, where potential superfluid phases are expected to arise.

In a similar double-layer system involving two coupled moir\'e bilayers the correlated excitonic insulating phases were observed at fractional fillings \cite{Zeng2023exciton}. Here, one bilayer is charge-neutral, while the other is doped to a generalized Wigner crystal state at fractional filling. A perpendicular electric field transfers electrons from the Wigner crystal to the charge-neutral bilayer, and the trasferred electrons bind to the holes left behind, forming long-lived interlayer excitons. Due to intralayer electron-electron repulsion, excitons are expected to only hop or diffuse along channels defined by the Wigner crystal formed by electrons. Because of this, excitons are expected to break translation symmetry, and the resulting state can be understood as an exciton density wave.


\subsection{Hubbard-like Exciton-Exciton Interactions in Moiré Superlattices}

Recent experiments on  WSe$_2$/WS$_2$ moiré heterobilayers revealed clear signatures of strong exciton-exciton interactions in a Bose-Hubbard-like system governed by an effective onsite repulsion dramatically enhanced due to the dipolar nature of the interlayer excitons~\cite{Erkensten2021exciton,Park2023dipole,Deng2025frozen}.  Under increasing power, the PL spectrum evolved into a ladder of discrete emission peaks, each corresponding to successive exciton occupation numbers per moiré site. The lowest-energy peak, IX$_1$, appears at energy $E_X$ and is attributed to single exciton occupancy. As the density increases, a second peak, IX$_2$, emerges at a higher energy, quantifying the exciton-exciton repulsion $U_{\mathrm{ex-ex}}$ associated with double occupancy. Higher energy peaks (IX$_3$, IX$_4$) are observed at still larger powers, and are interpreted as signatures of triple occupancy or population of higher moiré orbitals. The appearance of the peaks follows a very intuitive equation,
\begin{equation}
 N\omega_p=N\omega_X+U_{\mathrm{ex-ex}}\frac{N(N-1)}{2}, 
\end{equation}
which indicates that the energy $\omega_p$  of $N$ emitted photons matches the energy of $N$ interacting excitons~\cite{Camacho2022moire}.  This multi-exciton resonance leads to a discrete lobular pattern resembling the superfluid-Mott transition observed, for instance, in quantum gases~\cite{Greiner2002quantum}. In this case, the lobes follow a different condition than their equilibrium counterpart reflecting the underlying non-equilibrium character of the system~\cite{Camacho2022moire}.
The non-equilibrium dynamics manifests in phenomena such as polarization switching and Pauli blocking near the Mott regime, as recently reported in similar moiré systems~\cite{Kim2024correlation}.

\begin{figure}[b]
\includegraphics[width=.85\columnwidth]{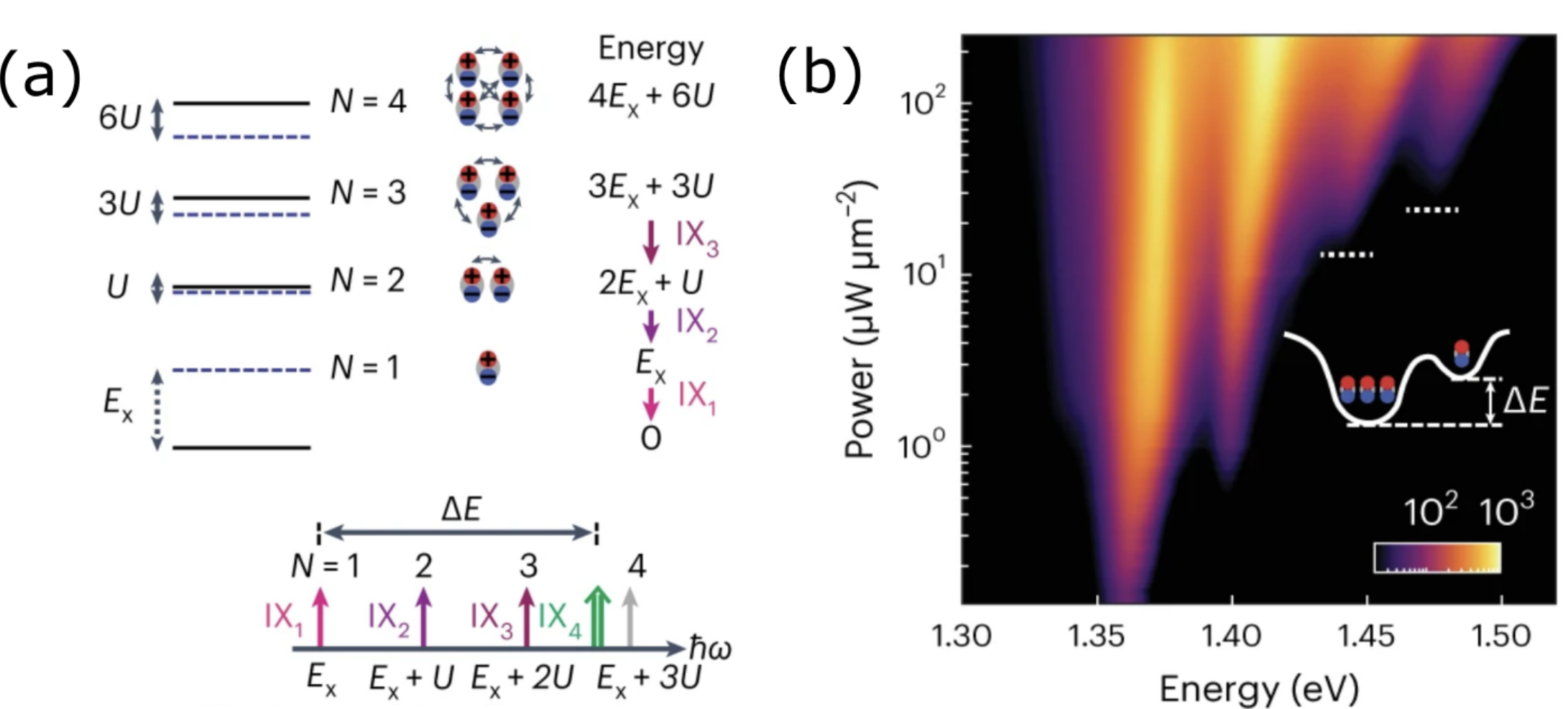}
\caption{\textbf{Observation of a dipolar exciton ladder in a WSe$_2$/WS$_2$ moiré lattice.} (a) Energy level diagram for a bosonic Hubbard ladder, showing successive exciton states in a single moiré trap separated by an onsite interaction energy $U$. (b) PL spectrum as a function of excitation power reveals distinct peaks corresponding to up to four excitons localised at a single site. Inset: illustration of moiré confinement potential and dipole repulsion. Figure adapted from {\emph{Park et al.}~\cite{Park2023dipole}.}}
\label{fig:DipolarLadders}
\end{figure}

Figure~\ref{fig:DipolarLadders} summarizes these key observations: panel (a) illustrates the energy ladder expected from on-site interactions in a single moiré trap, while panel (b) shows the measured PL spectrum as a function of power. The approximately equidistant energy levels support a bosonic ladder model with strong on-site repulsion. The co-circular polarization of higher-lying peaks further suggests the involvement of additional orbital states~\cite{Park2023dipole}. 

In the following sections, we discuss three intimately connected developments in the study of moiré quantum materials: the use of excitons as probes of strongly correlated electronic phases, the physics of Bose--Fermi mixtures, and the formation of exciton polarons. While we present these topics separately for clarity, the boundaries between them are fluid, as they all describe different aspects of the same underlying physics, namely, the interplay between mobile charge carriers and excitonic quasiparticles in population-imbalanced/balanced regimes.

In many cases, excitons act as localised, polarisable bosons embedded in a Fermi sea of electrons or holes, forming Bose-Fermi mixtures where the degree of imbalance and interaction strength dictate the  behavior. This framework naturally leads to the exciton-polaron picture, where fermionic carriers dress the excitons, giving rise to polaron branches. These renormalized excitonic features not only reveal the presence of Fermi seas, but also encode signatures of their correlations, such as Mott physics, charge order, or Fermi surface reconstruction. 

Conversely, when the exciton and carrier populations are more balanced, the many-body dynamics becomes inherently more complex, and the system may exhibit hybrid collective behavior, such as bound states, coherence effects, or even condensate formation. Thus, the distinction between probe, quasiparticle, and collective many-body phases blurs, revealing a unified landscape of boson-fermion interactions in moiré systems.

\subsection{Excitonic Probes of Correlated States in Moiré Superlattices}

Excitons in van der Waals materials provide a powerful mean to access and probe exotic quantum many-body phases ~\cite{Shimazaki2021Optical,Zeng2022strong,Salvador2022Optical,Julku2024exciton,Kiper2025confined,Huang2025collective}. That is, excitons can not only form correlated phases of matter but can be exploited as non-invasive optical sensors able to map complex phase diagrams in quantum matter.

\begin{figure}[b]
\includegraphics[width=.85\columnwidth]{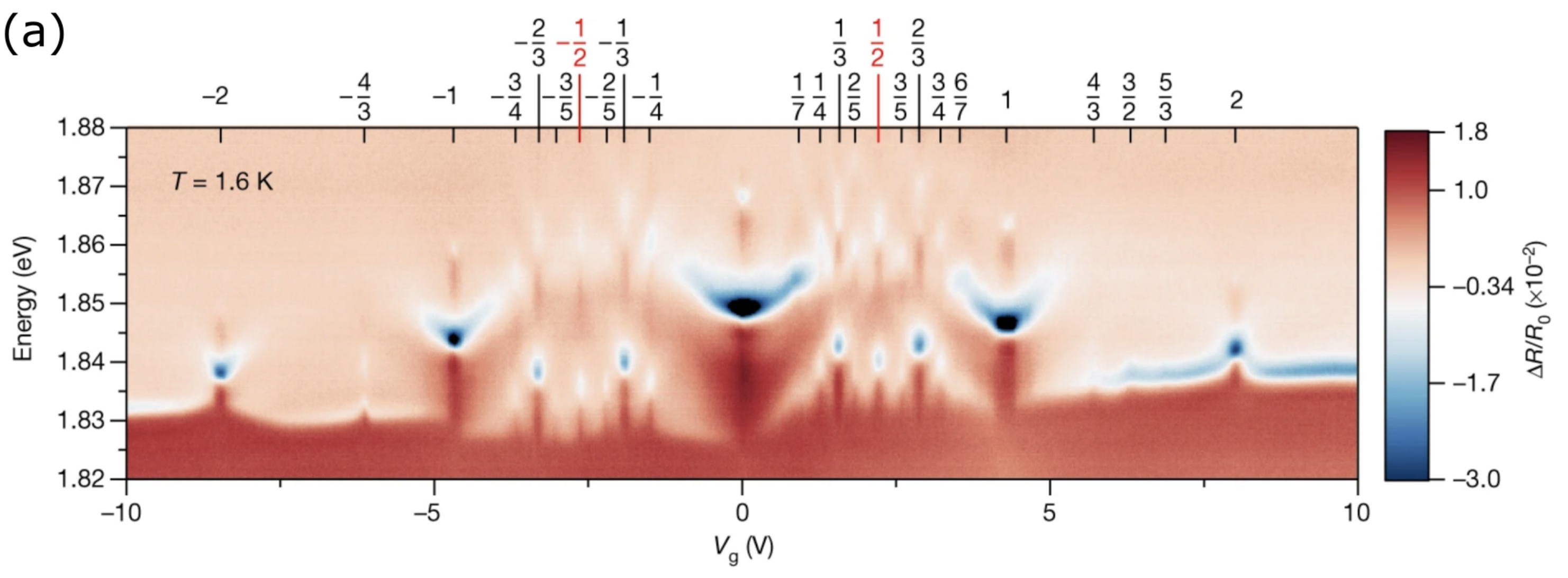} 
\caption{\textbf{Optical detection of correlated insulating states at fractional fillings in a moiré superlattice.}
(a) Gate-dependent reflection contrast $\Delta R/R_0$ measured via a WSe$_2$ excitonic sensor placed near a WSe$_2$/WS$_2$ moiré superlattice. The 2$s$ exciton resonance in the WSe$_2$ sensor exhibits a series of blueshifts and enhanced spectral weight at specific gate voltages $V_g$, indicating the opening of charge gaps in the adjacent moiré system. The upper axis shows the corresponding filling factors $\nu$ of the moiré lattice, with several correlated insulating states observed at fractional values including $\nu = 1/3$, $2/3$, $1/2$, $2/5$, and beyond. Figure adapted from \emph{Xu et al.}~\cite{Xu2020correlated}.
}
\label{fig:Fractional_probe}
\end{figure}

One of the first proof of concept of sensing with excitons were Rydberg excitons. Rydberg excitons are particularly attractive in view of their sensitivity to dielectric screening due to their spatial extent and weaker binding energies. This dielectric sensitivity was employed in Ref.~\cite{Xu2020correlated} to detect a cascade of correlated insulating states in WSe$_2$/WS$_2$ moiré superlattices. By placing a monolayer WSe$_2$ in close proximity to an heterobilayer and optically monitoring its $2s$ exciton resonance, they observed discrete shifts in exciton energy and oscillator strength as a function of charge filling in the moiré lattice. This shift marked the onset of correlated insulated phases at fractional fillings, revealing several of such states without requiring electric transport measurements. This is illustrated in  Fig.~\ref{fig:Fractional_probe}, where the reflection contrast map reveals sharp, symmetric jumps in the $2s$ exciton energy centered around half-filling, which vanish at elevated temperatures—confirming their many-body origin. 


Ref.~\cite{Shimazaki2020strongly} demonstrated that hybridized excitons—formed in a MoSe$_2$/hBN/MoSe$_2$ moiré heterostructure optically probe a Mott-like incompressible electron state at half-filling. By tracking shifts in the exciton-polaron resonance as a function of gate voltages, they revealed a correlated electron phases and strong pseudospin paramagnetism. 

In a related effort, Ref.~\cite{Regan2020mott} demonstrated electrically tunable Wigner crystals and generalized Mott states in TMD moiré heterostructures. Their studies employed capacitance measurements and optical probes to identify incompressible phases at fractional fillings, supporting the presence of strong electron-electron correlations. The ability to resolve these states optically, through blueshifts in exciton energy and modified oscillator strengths, underscores the power of excitonic probes in revealing the fermionic landscape. These findings emphasize the feasibility of stabilizing strongly interacting fermionic phases within a moiré potential.
Excitons have also been used to demonstrate  a crystalline order even in the absence of a moiré superlattice, such as  Wigner crystallization~\cite{Zhou2021bilayer}, and to test signatures of fractional electronic filling~\cite{Liu2021excitonic}. The photoluminescence of interlayer excitons in  WSe$_2$/WS$_2$ moiré superlattices has also been used to probe correlated electron states by indicating the filling-dependent energy shifts, intensity enhancement, and valley polarization effects that signal interactions between excitons and Mott or Wigner-type insulating backgrounds \cite{Miao2021strong}.
This all-optical approach provides a powerful means of charting the correlated phase diagram of moiré materials, even in the absence of electrical contacts.

\subsection{Bose-Fermi Mixtures in Moiré Heterostructures}

Excitons and charge carriers can coexist, as they can be injected via different mechanisms—optical excitation and electrostatic gating, respectively—leading to the simultaneous presence of bosonic and fermionic (quasi-)particles. This feature opens the door to realizing and exploring Bose-Fermi Hubbard models in scalable solid-state platforms. In such systems, excitons may be intralayer, interlayer, or hybrid, while the fermions can consist of either electrons or holes. The interplay between these species gives rise to a rich phase diagram governed by the relative densities and interactions of the components.

The moiré superlattice not only localises these quasiparticles but also enhances and tunes their mutual interactions. Strong Coulomb repulsion can arise between all combinations of particles, including exciton-exciton ($U_{\textrm{ex-ex}}$), electron-electron ($U_{\textrm{e-e}}$), hole-hole ($U_{\textrm{h-h}}$), exciton-electron ($U_{\textrm{e-ex}}$), and exciton-hole ($U_{\textrm{h-ex}}$) interactions~\cite{Gao2024excitonic,Lian2024valley}. The resulting energy scales can be externally controlled by adjusting the dielectric environment, gate voltages and optical excitation, providing versatile handles to access and probe different many-body regimes.

\begin{figure}[b]
\includegraphics[width=.85\columnwidth]{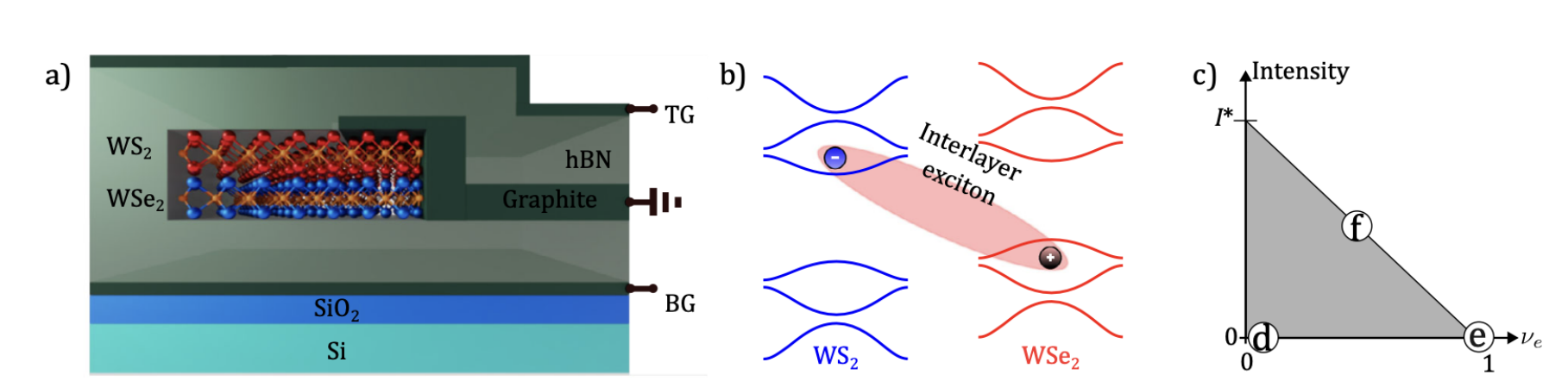}
\caption{Experimental realization of a Bose-Fermi Hubbard system in a WSe$_2$/WS$_2$ moiré heterobilayer. (a) The dual-gate TMD heterostructure enables the independent control of the electron and exciton densities via gating and optical excitation intensity, respectively. (b) Band alignment and schematic of the relevant interlayer exciton. (c) Typical phase diagram accessible for a Bose-Fermi Hubbard system which can range from Bose/Fermi-polarons to strongly correlated mixtures. The optical pumping intensity $I$ controls the exciton (bosonic) density, while a gate voltage controls the electronic (fermionic) filling factor $\nu_e$. Figure  (adapted from \emph{Gao et al.}~\cite{Gao2024excitonic}) }
\label{fig:Gao_BoseFermi}
\end{figure}

Recent experiments have demonstrated the realization of Bose-Fermi Hubbard physics in moiré superlattices \cite{Gao2024excitonic,Lian2024valley,Mhenni2024gateArxiv}. A particularly attractive feature of these systems is the independent tunability of bosonic and fermionic populations. As shown in Fig.~\ref{fig:Gao_BoseFermi} for a WSe$_2$/WS$_2$ moir\'e heterobilayer, the optical pumping intensity $I$ governs the exciton density, while a gate voltage $V_g$ controls the electronic filling factor $\nu_e$ (or $\nu_h$ for hole doping). This dual control enables access to distinct interaction regimes, from weakly interacting Bose-Fermi mixtures to regimes dominated by strong on-site repulsion and lattice commensuration. Experimental observables such as exciton photoluminescence shifts and reflectivity contrast offer sensitive probes of these interaction-driven transitions.

Although to some extent all experiments with moiré electrons in which excitons are used as probes can be regarded as Bose-Fermi systems, a distinction can be made in terms of the filling factors. In the very low excitonic density regime, the moiré excitons can be regarded as impurities in the medium and one can understand the highly population-imbalanced Bose-Fermi mixture in terms of polarons~\cite{Julku2024exciton}. However, the realm of the moiré Bose-Fermi-Hubbard models extends far beyond the impurity limit.

A realization of Bose-Fermi lattice physics  in Ref.~\cite{Zeng2023exciton}, reported the formation of exciton density waves in Coulomb-coupled dual moiré lattices WS$_2$/WSe$_2$/WS$_2$ multilayers. In this system, excitons arise as interlayer bound states between electrons and holes localised in two spatially separated moiré superlattices. At fractional total fillings $\nu = 1/3$, $2/3$, $4/3$, and $5/3$, correlated insulating states were observed, which were attributed to the formation of exciton density waves, bosonic states whose density modulation spontaneously breaks translational symmetry.

The interplay between charge carriers (fermions) and excitons (bosons) in the strongly correlated regime was recently explored in Ref.~\cite{Gao2024excitonic} In a bilayer of WS$_2$/WSe$_2$ the fermionic (electrons) and bosonic densities (interlayer excitons) were tuned independently via gating and optical pumping, respectively. The optical response revealed several regimes of the Hubbard model, from a dilute bosonic gas to strongly interacting phases with electron-exciton and exciton-exciton double occupancies inducing new peaks in the PL spectrum. The incompressibility of the excitonic states close to integer filling is signaled by an energy gap in the photoluminescence, together with a significant suppression of the diffusion. In a recent experiment, a giant enhancement of exciton diffusion was observed in a Bose-Fermi mixture where charge doping was near the Mott insulator phases, increasing the excitonic diffusion by three orders of magnitude~\cite{Upadhyay2025,Yan2024anomalouslyArxiv,Pichler2025purelyArxiv}.

The observation of several insulating states at fractional fillings  in WSe$_2$/WS$_2$ bilayers \cite{Xu2020correlated} (shown in Fig.~\ref{fig:Fractional_probe}) remarks the capability of moiré lattices to stabilize interaction-driven phases. In Ref.~\cite{Tan2025enhanced}, localised excitons coexist with itinerant charge carriers within the the same moiré lattice, where excitons serve as probes to measure  fermionic incompressibility with bosonic coherence. In these Bose-Fermi Hubbard systems, the role of the excitons can indeed be regarded as impurities. Bose-Fermi mixtures also allow the realization of intercell charged moiré exciton complexes \cite{Chen2023excitonicREVIEW,Barman2023interplay,Wang2023intercell}.

In general, moiré heterostructures offer a unique opportunity to emulate multi-component Bose-Fermi mixtures, where degenerate excitons interact with flat-band-confined electrons or holes. The tunability of the moiré potential and the control over interlayer coupling allow for the engineering of interaction strengths and effective lattice geometries. This includes regimes where hybrid Bose-Fermi Mott states, phase-separated mixtures, or coherent composite states may arise. Theoretical proposals suggest that such systems can host rich many-body phenomena including polaron condensation, exciton-mediated superconductivity, and exotic symmetry-broken states, positioning moiré materials as a highly versatile platform for quantum simulation of complex Hubbard models.

\subsection{Moiré Exciton Polarons.}

The polaron, a quasiparticle formed by an impurity dressed by its environment was first introduced by Landau and Pekar ~\cite{landau1948effective,PEKAR1958} to describe an electron coupled to lattice vibrations in polar crystals. This concept has become a powerful tool to describe many systems in  condensed matter physics, atomic physics, and even nuclear matter~\cite{2025arXiv250109618M,Grusdt_2025,Paredes2024}.  

In recent years, polaron physics has renewed interest in view of experiments with ultracold gases~\cite{2025arXiv250109618M,Grusdt_2025} that challenge the original picture of Landau and Pekar. In this context, it has been already suggested that lattice polarons ~\cite{Amelio2024,amelio2024polaron,Ding2024,Santiago2024lattice} exhibit contrasting behaviour compared to polaron in homogeneous environments~\cite{jorgensen2016observation,hu2016bose,yan2020bose,skou2021non}, and that polaron dressing can break the formation of molecular (exciton-like) states~\cite{Ardila2025}. 

The arrival of TMD's, unveiled new classes of Fermi/trion polarons~\cite{Efimkin2017,Efimkin2021,Tiene2020,Tiene2022,Muir2022,Wilson2021,2025arXiv250109618M,Huang2023quantum} and Fermi polaron-polaritons  with novel linear and non-linear properties~\cite{Sidler2017,tan2020interacting,gu2024giant,rana2021exciton,Emmanuele2020,Li2021b,Kumar2023,bastarrachea2021polaritons,Bastarrachea2021,Bastarrachea2024} with a tunability similar to the ultracold gases. As we have discussed, highly population-imbalanced exciton-electron/holes mixtures can be understood in terms of polarons. In contrast to conventional 2D semiconductors, moiré polarons are regarded as lattice polarons, where either the moiré superlattice or the modulation of a Fermi-Bose sea confines the excitons to the moiré traps~\cite{Mazza2022,Campbell2022exciton,Evrard2025ac, Cho2024moire,Julku2024exciton}. 


Recent ac Stark spectroscopy experiments on electron-doped MoSe$_2$/WS$_2$ heterostructures have revealed qualitative deviations from the standard Fermi-polaron picture~\cite{Evrard2025ac}. In contrast to monolayer behavior, attractive polarons in a deep moiré lattice show density-independent light shifts and saturable responses characteristic of localised, non-interacting emitters. These observations suggest that, rather than forming extended quasiparticles hybridized with a broad Fermi sea, moiré exciton polarons are localised at individual moiré sites, exhibiting suppressed intersite hopping and negligible mutual interactions.

 Moreover, by controlling the twist angle of nearby hBN layers, the polaron energy landscape can be further tuned via long-range dipolar interactions and dielectric screening, as recently demonstrated in twisted MoSe$_2$/hBN/MoSe$_2$ trilayers~\cite{Cho2024moire}.

Recent works have explored how coupling to either phonons or charge carriers modifies the nature of moiré excitons, leading to polaronic effects. Ref.~\cite{Knorr2024} developed a microscopic theory showing that the exciton-phonon coupling 
in MoSe$_2$/WSe$_2$ leads to the formation of moiré polarons, which exhibit an enhanced effective mass, temperature-dependent band flattening, and a twist-angle dependent suppression of hopping.

\subsection{Moiré Trions in TMD Heterobilayers.}
Recent experiments in transition metal dichalcogenide (TMD) moiré heterostructures have uncovered the formation of moiré-trapped trions, charged interlayer excitons localised by the periodic potential landscape in twisted bilayers, evidence of moir\'e trions were first observed by~\cite{Calman2020,Wang2021moire,Marcellina2021evidence,Dandu2022electrically,Zhao2024hybrid}. In Ref.~\cite{Wang2021moire}, in H-stacked MoSe$_2$/WSe$_2$ heterobilayers, the application of electrostatic doping enabled the filling of moiré potential minima with electrons or holes, resulting in the emergence of narrow photoluminescence (PL) peaks approximately 7\,meV below the neutral moiré exciton emission. These features correspond to positively and negatively charged moiré trions, denoted $M_T^+$ and $M_T^-$, and exhibit valley-dependent optical selection rules. The trions inherit valley Zeeman splitting and g-factors similar to their neutral exciton counterparts, supporting their assignment as moiré-confined species. Crucially, the polarization-resolved PL reveals a striking helicity reversal between $M_T^+$ and $M_T^-$, attributed to the competition between spin-conserving valley-flip and valley-conserving spin-flip relaxation channels during trion formation. 

Coulomb staircases have been observed in moiré superlattices as stepwise changes in the trion emission energy due to Coulomb interactions with carriers at nearest-neighbour moiré sites~\cite{Baek2021optical}.

The observation of trions localised in moiré traps introduces a new paradigm for engineering fermionic many-body states in optical lattices. Unlike excitons, trions are fermions, and their long-lived population (hundreds of nanoseconds) opens possibilities for correlated states with tunable spin, valley, and charge degrees of freedom. The polarization tunability of moiré trions under gate control, along with their sensitivity to electric fields via the Stark effect, enables dynamic control of optical emission properties in quantum emitter arrays. Furthermore, the relatively shallow confinement potential ($\sim$30\,meV) and small trion binding energies reflect the interplay between moiré length scales and trion Bohr radii, indicating a nontrivial regime for future theoretical modeling. These results establish moiré trions as optically addressable fermionic quasiparticles and pave the way toward hybrid excitonic circuits that combine bosonic and fermionic quantum degrees of freedom.

In twisted  MoSe$_2$ bilayers, the interplay between Coulomb interactions, interlayer hole tunneling, and the moiré potential landscape enabled a novel Feshbach  resonance mechanism for trions. In particular, Ref.~\cite{Schwartz2021} unveiled electrically tunable Feshbach resonances between excitons and holes residing in different layers allowing for tuning the exciton-hole scattering over a wide regime of interactions strengths \cite{Kuhlenkamp2022tunable}. The ability to control the exciton-fermion interactions in a moiré lattice opens pathways for engineering Bose-Fermi systems exploiting the analogue to cold atom resonances for quantum simulation purposes \cite{Zerba2024Realizing,Zerba2025Tuning}.  

\section{Moiré Exciton Polaritons: From Linear Regimes to Nonlinear and Topological Phenomena}\label{Sec: polaritons}

\begin{figure}[b]
\centering
\includegraphics[width=.5\columnwidth]{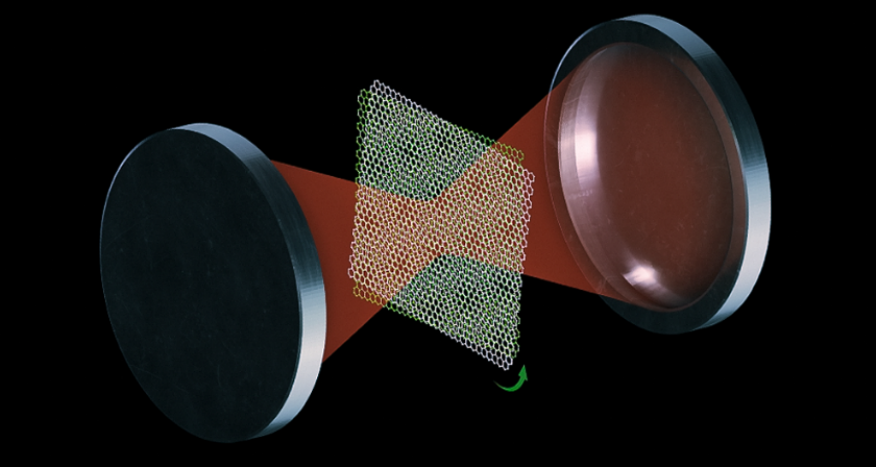}
\caption{Schematic of a moiré exciton polariton system. A twisted TMD heterobilayer is placed within a planar microcavity, enabling strong coupling between confined exciton states in the moiré potential and cavity photons. The hybridization yields moiré exciton polaritons—quantum fluids of light shaped by the moiré lattice geometry. Figure adapted from {\it Fitzgerald et al}.~\cite{Fitzgerald2022}}
\label{fig:MoireCavity}
\end{figure}

The advent of multilayered transition metal dichalcogenide (TMD) multilayers has enabled unprecedented control over light-matter interactions ~\cite{Dufferwiel2015,Lin2021twist,Turunen2022, Topp2021,Latini2019,König2023,Louca2023interspecies,Kang2023excitonREVIEW,Wang2024quantum,Moroni2025} and novel regimes of polariton interactions~\cite{Lin2024moire,Datta2022highly}. As we have mentioned over this Review,  moiré superlattices give rise to spatially modulated potentials for excitons, creating discrete energy levels and confined states. When embedded in optical microcavities, these excitons may couple to cavity photons, forming hybrid light-matter quasiparticles known as moiré exciton-polaritons, illustrated schematically in Fig.~\ref{fig:MoireCavity}.

In this section, we review the recent experimental and theoretical progress on moiré polaritons, which involve novel non-linear phenomena, the ability to engineer their dispersion and induce topological effects, as well as the potential realization of quantum many-body polaritonic phases.

\subsection{Experimental Realization and Nonlinear Optical Response}

Hybridized excitons in semiconductor bilayers hold particular potential for realizing nonlinear optical responses in condensed matter platforms. Because they inherit an appreciable oscillator strength from their intralayer component and a permanent dipole moment from their interlayer component, they can be expected to strongly couple to light and realize interacting polaritons. This is the case even in the absence of a moir\'e pattern, as TMD multilayers have been shown to host polaritons with potential for nonlinear optics \cite{Datta2022highly,Zhao2023exciton,Louca2023interspecies}. Under a moir\'e potential, polaritons can be further expected to exhibit enhanced nonlinearities, due to the moir\'e-induced exciton confinement.

Moiré exciton-polaritons, resulting from the strong coupling of moiré excitons with microcavity photons, were realized in a MoSe$_2$-WS$_2$ heterobilayer placed in a planar microcavity \cite{Zhang2021van}. A strong nonlinearity was found to arise as a consequence of the underlying dipolar character of hybrid excitons and their moir\'e induced localization.

\begin{figure}[b]
\centering
\includegraphics[width=.8\columnwidth]{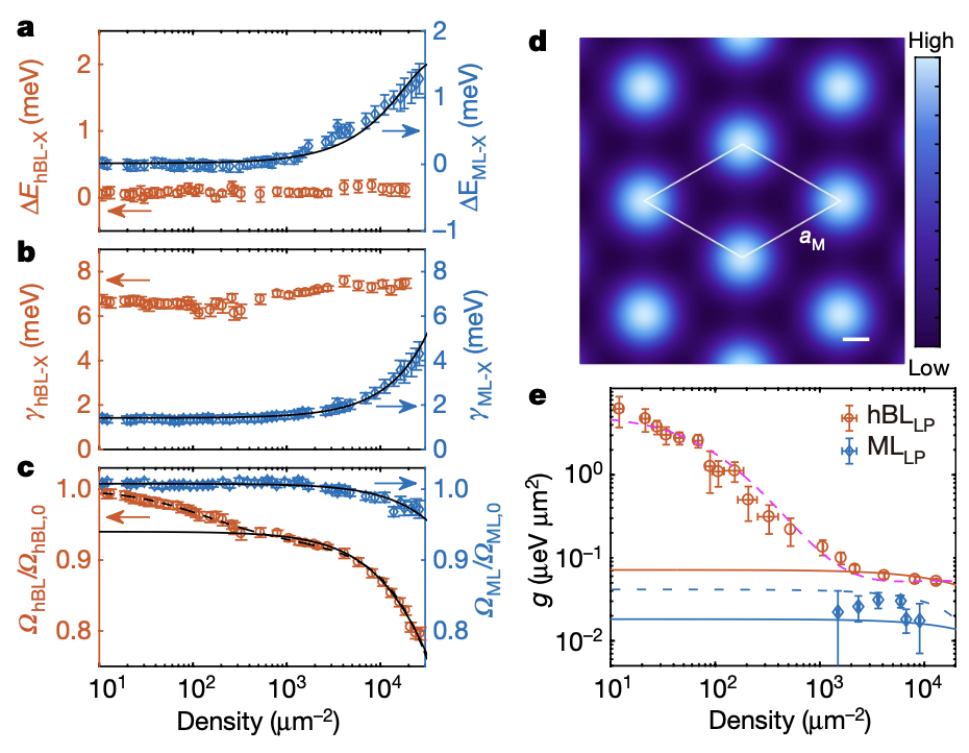}
\caption{Enhanced nonlinear optical response of moiré exciton polaritons. (a-c) Exciton energy shift $\Delta E$, linewidth $\gamma$, and normalized coupling strength $\Omega$ as function of excitation density for a moiré heterobilayer (hBL) and a MoSe$_2$ monolayer (ML), marked by red and blue circles, respectively. (d) Real-space map of the exciton localisation (interlayer component) in the moiré unit cell. (e) Nonlinear interaction coefficient $g$ extracted from the density-dependent blueshift of the lower polariton (LP), highlighting the enhanced nonlinearity of moiré polaritons. Figure adapted from \textit{et al.}~\cite{Zhang2021van}.}
\label{fig:ZhangEnhancedNL}
\end{figure}

Figure~\ref{fig:ZhangEnhancedNL} summarizes the optical response of the moir\'e polaritons in the heterobilayer (hBL), which contrasts with that of free 2D monolayer (ML) polaritons.
In Fig.~\ref{fig:ZhangEnhancedNL} (a)-(c), the quasiparticle properties of the polaritons (energy shift $\Delta E$, linewidth $\gamma$, and coupling strength $\Omega$) are shown as a function of density. The coupling strength for the moiré polaritons (red circles) dramatically drops as the density of optical excitations is increased, while the energy shift and line-broadening remain negligible. Such signatures are expected from an exciton blockade effect, characteristic of localized, 0D excitons. In contrast, monolayer polaritons (blue circles) retain their light-matter coupling even at high densities. As shown in Fig.~\ref{fig:ZhangEnhancedNL} (e), the density-dependent nonlinear coefficient $g=|dE(n)/dn|$ indicates a significant larger nonlinearity for the moir\'e polaritons. The enhanced nonlinearity of moir\'e polaritons was attributed to the exciton blockade effect, arising due their zero-dimensional character \cite{Zhang2021van}.



\subsection{Theory: Moiré exciton-polaritons}

Engineering the polariton dispersion of moiré polaritons via the twist angle was explored theoretically in Ref.~\cite{Fitzgerald2022}, developing a comprehensive microscopic theory to understand the tunability of the moiré polariton landscape. 
The energy detuning between moir\'e excitons and the cavity photon, as well as the number and dispersion of polariton branches is very sensitive to the moiré period.
At small twist angles ($\sim$1$^\circ$), multiple flat excitonic minibands with a strong oscillatory strength arise, leading to several polariton branches. At larger angles ($\sim$3$^\circ$), the oscillator strength of the moir\'e excitons is redistributed, transferring most of it to the lowest-energy exciton. Although moir\'e excitons exhibit flat bands (or a large effective mass) and are highly localized at low angles, the cavity photon is delocalized in the transverse plane, and it is expected to partially inherit its delocalized nature to the polaritons. Increasing the photonic contribution to the polaritons by tuning the cavity length is expected to reduce the effective mass of the polaritons, making them delocalized over many moir\'e cells, even for a small photonic component. Therefore, moiré polaritons are in general expected to exhibit strong nonlinearities and to partially inherit the twist-angle tunability of moir\'e excitons.

Topological transport effects of moiré polaritons in TMDs were theoretically explored in Ref.~\cite{Yu2020electrically} (see also Ref. \cite{Gutierrez2018polariton,Phuong2023electron}). This study exploited the interlayer character of moiré excitons to control properties of the polaritons via electrical means. The authors predicted several topological transport phenomena including electrically tunable valley and polarization Hall effects. The spin-valley locking together with the spin texture can lead to non-trivial Berry curvature.  By controlling the interlayer bias, the exciton-cavity detuning can be dynamically tuned, enabling real-time control of the polariton band topology and transport characteristics. These findings position moiré polaritons as a promising platform for electrically controlled topological photonic devices.

To explore the role of interactions beyond the perturbative regime, Refs.~\cite{Camacho2022moire,Camacho2022optical} proposed a driven-dissipative Bose-Hubbard model for moiré exciton polaritons. The model accounted for strong on-site exciton-exciton repulsion, saturable light-matter coupling, and coherent drive within a Lindblad master equation framework. The resulting phase diagram revealed bistable steady states and multiphoton resonances, and the formation of states with broken translational symmetry, features that are reminiscent of Mott physics and optical nonlinearities in Rydberg systems. When incoherent pumping was included, the system supported single- and multiphoton lasing regimes. These results suggest that moiré polaritons can realize strongly interacting photonic phases, enabling quantum simulations of lattice boson models in solid-state settings.


\section{Outlook}

Moiré excitons, their quantum many-body phases, and moiré exciton-polaritons are deeply interconnected, forming a unified framework for exploring strongly correlated and optically active quantum matter. At the core lies the ability of moiré patterns to confine excitons with lattice-scale control, enabling the engineering of discrete energy levels, optical selection rules, and spin-valley textures. These properties, in turn, provide a versatile platform for realizing and probing bosonic lattice models with tunable interactions, leading to the observation of quantum many-body phases such as Mott insulators, Wigner crystals, and superfluids. When embedded in optical cavities, these moiré excitons couple coherently to photons, giving rise to hybrid exciton-polariton states that inherit both the nonlinearity of the excitons and the delocalisation of the photons. As a result, moiré polaritons naturally extend the concept of strongly correlated excitonic phases into the driven-dissipative and topological photonic domains, offering new possibilities for quantum simulation, light-based computation, and the exploration of nonequilibrium many-body physics.

\textbf{Moiré Excitons.}
Moiré physics has unveiled new classes of excitons. The experimental realization of new families of optical excitations beyond the conventional Wannier-Mott framework, e.g., charge-transfer excitons and Rydberg moiré excitons, defy continuum theories and invoke new atomistic theoretical approaches. These excitons cannot longer be treated within descriptions based on center-of-mass confinement and instead demand multiscale modeling approaches that integrate electronic reconstruction, interlayer tunneling, and many-body interactions. 

The strong dependence of the underlying moiré superlattice on the twist angle, and the inherent valley optical properties of excitons, leads to the so-called 
\textit{valleytronics} and \textit{twistronics}, fields where the spatial modulation of optical selection rules within moiré lattices offers site-resolved valley control, enabling helicity-patterned emission and electrically switchable valley pseudospins. Combined with interlayer hybridization and twist-angle engineering, this creates a versatile platform for potential on-chip valleytronic architectures where spin-valley information can be stored, routed, or filtered with nanometric precision. At the same time, the twist angle continues to serve as a powerful design parameter, controlling miniband formation, exciton localisation, and even effective dimensionality.

Quantum-dot-like emitters in moiré superlattices represent another breakthrough, effectively creating periodic arrays of quantum light sources with built-in helicity and tunable confinement. The ability to selectively address excitons localised at different moiré sites may enable scalable arrays of indistinguishable single-photon emitters, with potential applications in quantum communication and photonic quantum computing. Furthermore, controlled hybridization between interlayer and intralayer states provides a natural handle for manipulating these emitters' spectral positions and oscillator strengths via electric fields or twist angle.

Finally, moiré superlattices reshape not only electronic and excitonic landscapes but also the vibrational structure of the lattice itself. Moiré phonons—folded and symmetry-modified vibrational modes interact strongly with excitons and can serve as sensitive probes of local atomic registry. These phonon modes can participate in exciton relaxation, tunneling, and scattering processes, imprinting distinct signatures in resonant Raman and PL spectra.

\vspace{1em}

\textbf{Quantum Many-Body Phases with Moiré Excitons.}
Moiré heterostructures offer a platform for engineering quantum many-body phases with unprecedented control and scalability. Equipped with tunable parameters as lattice geometry, interactions and even the effective dimensionality, they have allowed for the realization of excitonic insulators, dipolar wave densities, and superfluids and have established moiré excitons as quantum simulators of lattice Hamiltonians. 

The ability to independently control the optical excitations and charge carriers has unfolded a panorama for the realization of uncharted phases of Bose-Fermi systems confined in optical lattices. This scenario extends from population-balanced scenarios, to impurity regimes where different many-body phases may emerge or excitons can be used as probe scheme to detect strongly correlated phases.

A particularly exciting frontier lies at the interface of polarons and Bose-Fermi physics. When a mobile charge carrier interacts with a dense excitonic background it forms a polaron—a quasiparticle resulting from dressing of excitonic excitations. In moiré lattices, the interplay between localisation, strong coupling, and band flattening allows access to previously unexplored polaronic regimes. These polarons can serve not only as fundamental quasiparticles but also as sensitive probes of the surrounding quantum fluid. Their spectral features such as linewidths, effective masses, and coherence encode the properties of the host medium \cite{Tan2025enhanced}, enabling momentum- and density-resolved spectroscopy of excitonic phases. 

The controlled coexistence of excitons and excess charge carriers also enables the realization of Bose-Fermi mixtures in engineered lattice geometries. This unlocks a zoo of quantum states: from phase-separated configurations to correlated Bose-Fermi complex many-body phases.

Many open questions remain to be addressed, for instance, the possibility to induce conventional and unconventional superconductivity via boson-mediated interactions, Feshbach physics, supersolid phases, and the realization of Bose-Fermi Hubbard systems in unexplored regimes.

\vspace{1em}

\textbf{Moiré Exciton-Polaritons}

Moiré exciton-polaritons merge the strong light-matter coupling of TMD heterostructures with the periodic confinement of moiré lattices, enabling a new regime of nonlinear and quantum photonics \cite{Du2024nonlinearREVIEW}. By engineering the excitonic landscape via twist angle, stacking, or pressure, it becomes possible to tailor polariton dispersion, localisation, and oscillator strength at the single-unit-cell level. This flexibility has led to demonstrations of enhanced nonlinearities, such as interaction-induced blueshifts and saturation effects, with interaction strengths orders of magnitude larger than in conventional monolayer systems. 

The physics of moiré polaritons is intrinsically interesting even within the linear regime, as they can potentially realize topological phases. The spin-valley physics of TMD excitons transfers a geometric structure to polaritons, leading to nontrivial Berry curvatures and valley-selective transport. Thus, pending experimental realization of electrically tunable valley Hall and polarization Hall effects as well as chiral polariton phases.  Topological polaritons are promising platforms for new opto-electronic devices, valley-selective routes, and topological lasers.

The interacting character of moiré polaritons places them at a rich intersection between strong light-matter physics and strong dipole-dipole interactions, together with their driven-dissipative nature. This opens up the door for the realization of dipolar resonances, quantum bistabilities, polaritonic Mott phases, non-equilibrium phase transitions and multi-photon lasers, among others.

From a quantum optics perspective, moiré exciton-polaritons offer a compelling route toward scalable platforms for coherent light–matter interfaces. The ability to localise polaritons in moiré-confined quantum wells enables site-selective addressing and manipulation of individual quantum states of light. Recent advances in resolving the coherence and interference patterns of single moiré excitons \cite{Durmus2023prolonged,Wang2024quantum,Tan2025enhanced} mark a critical step toward realizing arrays of strongly coupled, cavity-enhanced quantum emitters.

\vspace{1em}

\textbf{Quantum non-linear optics and transport}

A central goal in the field of strong light–matter coupling with two-dimensional (2D) materials is the realization of polariton blockade, a phenomenon that could unlock single-photon nonlinearities and all-optical switching at the quantum level \cite{Du2024nonlinearREVIEW}. However, achieving this requires a deeper understanding of the complex microscopic interactions governing exciton-polaritons, particularly in correlated materials where excitons, photons, phonons, and magnons are intricately coupled. Moiré lattices may introduce novel degrees of freedom that can help fine-tuning interactions at different levels and provide an elegant path towards quantum-based all-optical devices.
In terms of quantum information and communication, of growing interest are polariton Bose–Einstein condensates formed in antiferromagnetic crystals, where the inherently high-frequency magnons present exciting possibilities for quantum transduction. Here, moiré heterostructures may offer a fertile ground for exploring novel collective quantum phenomena arising from the interplay of light and spin excitations. 

Exciton-polaritons represent a unified opto-electronic platform that naturally integrates traditionally different processes: transport of charge carriers and light. The emerging fields of trion and Fermi polaron resonances in moiré heterostructures, as well their polariton counterparts in optical cavities, will be a fascinating avenue to bridge these separate experimental regimes, whereby the coupling to photons can enhance both ballistic and coherent transport and mitigate the effects of local disorder. This opens pathways toward a polariton-mediated transport regime with the long-term prospect of enabling unconventional mechanisms for superconductivity.

A major challenge in quantum light generation and transport is overcoming decoherence, which fundamentally limits the performance and fidelity of quantum systems. Enhancing the coherence times of excitons and spin states requires minimizing their interactions with the surrounding environment. Alternative strategies with respect to lowering temperature to extreme values may consist in the precise stacking of 2D heterostructures and advanced material encapsulation techniques. 

Understanding and controlling decoherence processes depends critically on advanced experimental techniques capable of resolving ultrafast dynamics. Since the advent of 2D materials, out-of-equilibrium spectroscopies have become essential tools for probing exciton formation, relaxation, and recombination processes across timescales ranging from tens of femtoseconds to nanoseconds. Among these, ultrafast multidimensional coherent spectroscopy—particularly two-dimensional electronic spectroscopy (2DES)—has proven uniquely powerful for unraveling many-body interactions and coherent couplings in TMD monolayers. Time-resolved tracking of the evolution of the density matrix in complex quantum materials, even under fast decoherence on the order of 100 femtoseconds will serve as a critical bridge between fundamental studies of coherence and the practical goal of engineering long-lived quantum states for integrated photonic applications.

Looking ahead, a critical step will be the seamless integration of single-photon emitters based on 2D materials and moiré heterostructures into on-chip photonic architectures \cite{Chen2025directional}. Embedding these quantum light sources into waveguides, optical resonators, and nanocavities will be essential for building compact, scalable quantum photonic circuits.

\section*{Acknowledgments}
A.C-G. acknowledges financial support from UNAM DGAPA PAPIIT Grant No. IA101325, Project SECIHTI (formerly CONAHCYT) No. CBF2023-2024-1765 and PIIF25. S.A.H.\ acknowledges financial support from SECIHTI (formerly CONAHCYT) Project No. 1564464.
G.P. acknowledges financial support from UNAM DGAPA PAPIIT Grant No. IN104325, Projects SECIHTI (formerly CONAHCYT) Nos. 1564464 and 1098652 and PIIF25. D.A.R.-T.\ acknowledges financial support from UNAM DGAPA PAPIIT Grant No.\ IN114125.

\bibliographystyle{unsrtnat}

\bibliography{references}

\end{document}